\documentclass[12pt]{article}
\pdfoutput=1
\pdfpageattr{/Group << /S /Transparency /I true /CS /DeviceRGB>>} 
\usepackage{graphicx}
\usepackage{amsmath}
\usepackage{amsfonts}
\usepackage{amssymb}
\usepackage{setspace}
\usepackage{subcaption}
\usepackage{caption}
\usepackage[hidelinks]{hyperref}
\usepackage{mathrsfs}
\usepackage{enumerate}
\usepackage[utf8]{inputenc}
\numberwithin{equation}{section}
\newcommand{\gsim}{\lower.7ex\hbox{$\;\stackrel{\textstyle>}{\sim}\;$}}
\newcommand{\lsim}{\lower.7ex\hbox{$\;\stackrel{\textstyle<}{\sim}\;$}}

\makeatletter
\newsavebox\myboxA
\newsavebox\myboxB
\newlength\mylenA

\newcommand*\xoverline[2][0.75]{%
    \sbox{\myboxA}{$\m@th#2$}%
    \setbox\myboxB\null
    \ht\myboxB=\ht\myboxA%
    \dp\myboxB=\dp\myboxA%
    \wd\myboxB=#1\wd\myboxA
    \sbox\myboxB{$\m@th\overline{\copy\myboxB}$}
    \setlength\mylenA{\the\wd\myboxA}
    \addtolength\mylenA{-\the\wd\myboxB}%
    \ifdim\wd\myboxB<\wd\myboxA%
       \rlap{\hskip 0.5\mylenA\usebox\myboxB}{\usebox\myboxA}%
    \else
        \hskip -0.5\mylenA\rlap{\usebox\myboxA}{\hskip 0.5\mylenA\usebox\myboxB}%
    \fi}
\makeatother
\newcommand{\be}{\begin{equation}}
\newcommand{\ee}{\end{equation}}
\newcommand{\bea}{\begin{eqnarray}}
\newcommand{\eea}{\end{eqnarray}}

\newcommand{\comment}[1]{}
\newcommand{\expect}[1]{\left\langle #1 \right\rangle}

\newcommand{\bsb}{\boldsymbol}

\def\ep{\epsilon}
\def\vep{\varepsilon}
\def\bvep{{\bsb \vep}}
\def\d{\partial}
\def\vphi{\varphi}
\def\x{{\bsb x}}
\def\out{{\langle {\rm out}|}}
\def\inn{{|{\rm in}\rangle}}
\def\O{\mathcal{O}}
\def\q{{\bsb q}}

\def\r{{\bsb r}}
\def\p{{\bsb p}}
\def\n{{\bsb n}}
\def\scri{{\mathscr I}}
\def\Q{{\mathscr Q}}
\def\innn{{\rm in}}
\def\A{{\mathcal{A}}}
\def\E{{\mathcal{E}}}
\def\S{{\mathcal{S}}}

\begin{document}

\vspace*{-1. cm}
\begin{center}

{\bf \Large Weinberg Soft Theorems \\[15pt]from Weinberg Adiabatic Modes}

\vskip 1cm
{\normalsize { Mehrdad Mirbabayi} and { Marko Simonovi\'c}}

\vskip 0.5cm

{\normalsize {\em Institute for Advanced Study, 1 Einstein Drive, Princeton, NJ 08540}}

\end{center}

\vspace{.8cm}

\hrule \vspace{0.3cm}
{\small  \noindent \textbf{Abstract} \\[0.3cm]
\noindent 
Soft theorems for the scattering of low energy photons and gravitons and cosmological consistency conditions on the squeezed-limit correlation functions are both understood to be consequences of invariance under large gauge transformations. We apply the same method used in cosmology -- based on the identification of an infinite set of ``adiabatic modes'' and the corresponding conserved currents -- to derive flat space soft theorems for electrodynamics and gravity. We discuss how the recent derivations based on the asymptotic symmetry groups (BMS) can be continued to a finite size sphere surrounding the scattering event, when the soft photon or graviton has a finite momentum. We give a finite distance derivation of the antipodal matching condition previously imposed between future and past null infinities, and explain why all but one radiative degrees of freedom decouple in the soft limit. In contrast to earlier works on BMS, we work with adiabatic modes which correspond to large gauge transformations that are $r$-dependent.


\vspace{0.3cm}
\hrule

\begin{flushleft}
\end{flushleft}

\vspace{-1cm}
\tableofcontents
\section{Introduction}

In recent years, there has been much progress in our understanding of the consistency conditions on squeezed cosmological correlation functions on the one hand, and soft theorems for the flat space scattering of low energy photons and gravitons on the other. The purpose of this paper is to make a connection between the two. 

The first examples of soft theorems in cosmology are known as Maldacena's consistency conditions \cite{Maldacena,Creminelli:2004yq}. They relate correlation functions with one spatial momentum sent to zero, to the correlation functions in the absence of the low momentum mode. They have been generalized \cite{Creminelli,Hinterbichler} using the notion of adiabatic modes introduced by Weinberg \cite{Weinberg_adia}, and have been understood \cite{Hinterbichler2} to follow as the Ward identities for spontaneously broken ``large'' diffeomorphisms. These are diffeomorphisms that do not vanish at infinity. Not all spontaneously broken large diffeomorphisms lead to consistency conditions for correlation functions of physical modes. Those which do are in one to one correspondence with adiabatic modes: large diffeomorphisms that can locally be mimicked by long wavelength physical perturbations.

In flat space scattering theory, Weinberg theorems for the emission of low energy (soft) photons/gravitons in the scattering of high energy (hard) particles have been known for a much longer time. They were originally derived using the universal cubic coupling of the gauge fields \cite{Weinberg}. Very recently, they have been understood to follow as the Ward identities for spontaneously broken asymptotic symmetry groups \cite{Strominger_graviton1,Strominger_graviton,Strominger_photon}. These are residual large gauge transformations that remain unfixed after local gauge-fixing, and in the case of gravity comprise the super-translation subgroup of BMS group acting at null infinity \cite{Bondi,Sachs}. The resulting Ward identities correspond to charge/energy conservation at every angle. Since the symmetries are spontaneously broken, there is a contribution from soft photons/gravitons as the associated Nambu-Goldstone modes. 

There is a clear similarity between the above identifications of cosmological consistency conditions and soft photon/graviton theorems as Ward identities. The major difference is that in cosmology there is a locally conserved current associated to each adiabatic mode, but the recently developed flat space derivations of Weinberg theorems are exclusively in terms of the asymptotic data. While this is most suitable in scattering theory, to understand the relation between the two, a derivation based on local conservation of a symmetry current would be very useful. 

Moreover, to make connection with physical observables, the asymptotic Ward identity must be expressible as the limit of a conservation law in terms of quantities at a finite distance. This can help understand the technical features and choices made in the asymptotic analysis of \cite{Strominger_graviton,Strominger_photon}. Some of the most notable ones among them are: (i) The reduction of polarization degrees of freedom of zero-frequency photons/gravitons to a single one. (ii) The qualitative difference between $4d$, where the long distance behavior of large gauge transformations and radiation agree, and higher spacetime dimensions, where they don't. This is in contrast to the original derivation of soft theorems which applies universally to all $d\geq 4$. (iii) The ``antipodal'' matching condition between symmetry groups at future and past null infinities, BMS$^+$ and BMS$^-$. Antipodal matching is the statement that the transformation at any direction on the sphere at future infinity must match the opposite direction on the sphere at past infinity. This is the only choice compatible with the Lorentz symmetry, and it is intuitive given that free scattering states map antipodal points to each other. However, it is unclear how such a discontinuous choice can arise from a finite distance point of view.

An interesting proposal in this direction has been made in \cite{Susskind} where an infinite set of conserved charges (one in each direction) were obtained in Maxwell theory by considering a simple example. Conserved quantities are defined on a large sphere surrounding a neutral particle which decays into charged jets. These nicely match the conserved charges at future null infinity defined in \cite{Strominger_photon} and in the case of a scattering process with neutral ingoing particles provide a satisfactory continuation to a finite distance. However this proposal is incomplete: (i) Because of the above-mentioned antipodal matching the generalization to the case of charged ingoing states is nontrivial. (ii) A direct generalization when massive charges are included or when higher spacetime dimensions are considered fails to agree with the asymptotic charges.

We address the above questions in this paper. More specifically, we show that

\begin{itemize}

\item Adiabatic modes can be defined in electrodynamics and gravity in asymptotically flat spacetime. Weinberg soft theorems can be derived from the local conservation of the associated currents. 


\item There are infinitely many adiabatic modes but they all degenerate into the same leading soft theorem for scattering amplitudes. 

\item There are infinitely many constraints on local correlations or equivalently on the OPE coefficients of charged operators fusing into the gauge-fixed photon/graviton field and its derivatives. These are the flat-space analogs of the cosmological consistency conditions on squeezed correlation functions.

\item In QED an infinite set of conserved BMS charges can be obtained from the local conservation of the currents associated to the adiabatic modes.\footnote{To emphasize the similarity with the gravitational case we denote the asymptotic symmetry group of QED by BMS.} They arise as the $R\to \infty$ limit of conservation laws defined on a finite sphere of radius $R$. They apply when there are charged ingoing states, massive hard states, and at higher spacetime dimensions.

\item Naively, these conservation laws differ from the identities derived in \cite{Strominger_photon,Strominger_QED} in that the matching is regular rather than antipodal. However, once the contribution from the dressing field of the charged in- and out- states is taken into account, the antipodal matching between the rest of contributions can be derived.

\item As $R$ becomes large the external charged particles move almost radially and hence they decouple from the tangential components of the field strength $F_{ab}$. As a result all but one degree of freedom in the Maxwell field decouple in the soft limit.

\item There is a fundamental difference between the construction of asymptotic conservation laws based on adiabatic modes and that given in \cite{Strominger_photon}. Adiabatic modes grow with $r$ and hence have a different $r$-dependence compared to both radiation and the large gauge transformations of \cite{Strominger_photon}. At a fixed $r=R$ the two transformations can be matched and hence lead to the same asymptotic conservation laws. However, there is no longer any qualitative difference between four and higher spacetime dimensions as the symmetry transformations do not (and as we will argue need not) scale the same way as radiation, even in $4d$.

\end{itemize}

Normally, what results from symmetry principles is considered to be very robust. However, at first sight the residual large gauge transformations seem to be a deficiency of the gauge fixing procedure. So why do they lead to physically interesting results? Does the universality of physical results make the residual large gauge transformations inevitable? There are reasons to believe the answer is yes. Consider a gravitational wave with a very long wavelength $\lambda$. In any gauge-fixing scheme this corresponds to some nonzero metric fluctuations $h_{\mu\nu}$. But by the Equivalence Principle short-distance observers can only detect the wave at the order $\lambda^{-2} h_{\mu\nu}$ which vanishes as $\lambda \to \infty$. Hence, there is a nonzero perturbation with no effect on the local physics. As long as this is the case there should exist a symmetry even if local gauge freedom is fixed.\footnote{We thank Matias Zaldarriaga for a conversation on this point.}

\section{Adiabatic modes in electrodynamics} \label{sec:photon}

In this section we define adiabatic modes in electrodynamics by following as closely as possible the approach used in cosmology \cite{Weinberg_adia,Hinterbichler2}. We will derive symmetry currents and use their conservation to derive soft photon theorems for the $\S$-matrix. Then we discuss the implications for local correlation functions and OPE coefficients. Later in section \ref{sec:BMS} we will make a closer connection with the discussion of asymptotic symmetries. 

Let us illustrate the underlying idea through an analogy with the theory of nucleons and massless pions. This theory has a global axial symmetry that is spontaneously broken. It changes vacuum to a neighboring degenerate vacuum, or equivalently shifts the pion field uniformly $\pi \to \pi +c$; it also changes the phase of nucleon fields. The conservation of the axial current can be used to derive soft pion theorems \cite{pion}. The interaction of hard particles in the presence of a soft pion is equivalent to the interaction of hard particles in a slightly different vacuum, which is related to their interaction in the original vacuum by axial symmetry. This reasoning works because the uniform shift $\pi\to \pi+c$ is guaranteed to be locally mimicked by long wavelength perturbations. The equation of motion for the soft $\pi$ is a hyperbolic equation $(\d_0^2 - \d_i^2)\pi =0$. Once the constant shift $\delta\pi=c$ is deformed to be a mode of finite wavelength, the time derivative $\d_0^2\delta \pi$ will adjust itself to satisfy the equation of motion. 

In gauge theories, large gauge transformations play a similar role as the axial symmetry in the above example. They change the vacuum via generating infinite wavelength perturbations of the gauge field that do not vanish (neither oscillate) at infinity. The main difference is that since some of the equations are constraints, it is not guaranteed that the wavelength can be continued to a finite value. Hence, not all large gauge transformations can be locally mimicked by a soft photon (or graviton). In analogy with cosmology, we define adiabatic modes as the subclass of large gauge transformations for which this continuation is possible.

\subsection{Definition}

The Maxwell theory has a $U(1)$ gauge symmetry under which the fields transform as
\be
A_\mu \to A_\mu + \d_\mu \alpha,\qquad \phi \to e^{iQ\alpha} \phi
\ee
for any charged field $\phi$ of charge $Q$. To fix it we impose the temporal gauge condition\footnote{We work in mostly plus signature and do not distinguish between upper and lower spatial indices. We often use bold face variables to denote space vectors, e.g. $\q \cdot \x$, and normal variables for four-vectors, e.g. $q\cdot x$. Spatial gradient is denoted by $\nabla$.}
\be
A_0 =0.
\ee
This leaves time-independent gauge transformations with $\d_0\alpha(t,\x)=0$ unfixed. Such gauge transformations transform the vacuum into a state with nontrivial $A_\mu$ and hence, by virtue of being a symmetry, generate new solutions of the theory. 

These solutions are unphysical since they have zero frequency. However, there is a subclass of them that can be continued to finite frequency following Weinberg \cite{Weinberg_adia}. This is achieved by analyzing the homogeneous (source-free) part of the constraint equations and making sure they are not accidentally satisfied because frequency $\omega$ vanishes. This requirement forces $\alpha$ to be harmonic: The homogeneous constraint equation reads
\be\label{constraint}
\d_0\d_i A_i =0,
\ee
which is satisfied by any $A_i = \d_i \alpha(\x)$. However, to ensure that this solution can be continued to a finite frequency solution, we impose the stronger condition $\d_iA_i =0$  on adiabatic modes. It implies that 
\be\label{laplace}
\nabla^2\alpha(\x) =0.
\ee
Such gauge transformations are necessarily nontrivial at spatial infinity and are called large transformations. But they generate adiabatic modes since the dynamical source-free equation for $A_i$ is 
\be
(-\d_0^2 +\nabla^2)A_i-\d_i\d_jA_j=0
\ee
and there is no obstacle in continuing an infinite wavelength solution $A_i(\q =0)$ with $\d_jA_j = 0$ to finite wavelength. $\d_0^2A_i$ can adjust itself to cancel $\nabla^2 A_i$. 

Adiabatic modes can be organized in a Taylor expansion
\be\label{vep}
\alpha(\x) = \sum_{n=0}^\infty \frac{1}{(n+1)!} \bvep_{i_0\cdots i_n} \x^{i_0}\cdots \x^{i_n},
\ee
where $\bvep$ is maximally symmetric and repeated indices are summed over. Equation \eqref{laplace} implies that 
\be\label{trace}
\bvep _{ii i_2\cdots i_n} =0.
\ee
The conservation of the current associated to each adiabatic mode leads to a soft photon theorem. 

\subsection{Adiabatic modes as locally unobservable physical solutions}

Originally, Weinberg introduced adiabatic modes as a shortcut to find a universal solution in the theory of cosmological perturbations in the presence of multiple components. However, one can reverse the logic and find adiabatic modes (and hence new symmetries) by solving the source-free Maxwell equations and take the long wavelength limit. For instance, the lowest order term in \eqref{vep} arises from transverse plane wave solutions. More generally, consider the following Cauchy problem in the source-free Maxwell theory. At $t=0$ we take $A_i$ to be given by the following expansion around the origin
\be\label{An}
A_i(0,\x) =\sum_{n=0}^\infty \frac{1}{n!} a_{ii_1\cdots i_n} \x^{i_1}\cdots \x^{i_n},
\ee
where $a_{ii_1\cdots i_n}$ is symmetric in its last $n$ indices. Because of the constraint \eqref{constraint} $\d_i A_i$ is non-dynamical and we set it to zero, implying $a_{iii_2\cdots i_n} =0$. Otherwise the coefficients are arbitrary. In order to satisfy the boundary conditions at spatial infinity, this field is modified at distances $|\x| \sim \lambda$ and is no longer described by \eqref{An}. To complete the Cauchy data we assume
\be
\d_0 A_i(0,\x) =0.
\ee
Since $A_0=0$, the electric field vanishes at $t=0$. Now it is easy to identify the adiabatic modes \eqref{vep}. They correspond to configurations with zero magnetic field at $t=0$, with 
\be
a_{ii_1\cdots i_n} = \bvep_{ii_1\cdots i_n}.
\ee
As they evolve the electric and magnetic fields around the origin remain zero until the information about the deformation at distance $\lambda$ arrives:
\be
F_{\mu\nu}(t,\x) =0 \qquad \text{for}\quad |t|,|\x|\ll \lambda.
\ee
Any adiabatic mode can be obtained by superposing long wavelength transverse plane waves.\footnote{See \cite{double} for a discussion on this point.} In the limit $\lambda\to \infty$ we have
\be
\begin{split}
A_i(0,\q) =& \int d^3\x e^{-i\q\cdot\x}\sum_{n=0}^\infty \frac{1}{n!} \bvep_{ii_1\cdots i_n} \x^{i_1}\cdots \x^{i_n}
\\[10pt]
=& \sum_{n=0}\frac{i^n}{n!} \bvep_{ii_1\cdots i_n}\frac{\d^n}{\d \q_{i_1}\cdots \d\q_{i_n}}
(2\pi)^3\delta^3(\q).
\end{split}
\ee
Once the four-momentum of the photon $q^\mu$ is fixed the condition $\d_iA_i=0$ implies an extra transversality condition \cite{Hinterbichler2}
\be\label{trans}
\q_i \bvep_{i i_1\cdots i_n}=0.
\ee
In summary, there are admissible solutions of gauge-fixed Maxwell theory that are nontrivial but have no observable effect for $|t|,|\x|\ll \lambda$. Thus, in that spacetime region we have $A_\mu = \d_\mu \alpha$. Given that local gauge freedom is fixed, $\alpha$ is a large gauge-transformation which generates adiabatic modes.
\subsection{Noether current}

Having found a new symmetry of the gauge-fixed Maxwell theory, we next derive the Noether current for this symmetry $K^\mu$. This can be derived by noticing that the gauge fixed action is invariant under a gauge transformation with $\d_0\alpha = \nabla^2 \alpha =0$. The resulting Noether current is
\be\label{Jalpha}
K^\mu = \d_i\alpha F^{\mu i} + \alpha J^\mu,
\ee
whose conservation can be verified using Maxwell equation $\d_\nu F^{\mu\nu}= J^\mu$, and conservation of electric current $\d_\mu J^\mu =0$.\footnote{This current is conserved even if we relax the adiabaticity condition $\nabla^2 \alpha =0$. But we will only consider the subclass $\nabla^2 \alpha =0$. In a gauge which fully fixes local gauge degrees of freedom, such as the Coulomb gauge $\d_iA_i =0$, only adiabatic modes have a conserved current. A detailed derivation of adiabatic modes and their conserved current in Coulomb gauge is given in appendix \ref{app:Noether}.}

In the next few sections we discuss the implications of the conservation of this current for flat space scattering amplitudes as well as local correlation functions. We note in passing that by making a special choice of time-slicing and distinguishing time and space components the above symmetries can be naturally expressed as constraints on the wavefunction of the universe. As such they can be generalized to de Sitter space or cosmological backgrounds as in \cite{Pimentel,Hui}.

\subsection{Weinberg's theorem for soft photons}\label{sec:photonwein}

To derive Weinberg soft photon theorem we follow the standard derivation of soft pion theorems \cite{pion}. Current conservation implies
\be\label{dK2}
\int d^4x \ e^{-iq\cdot x} \d_\mu \out K^\mu(x) \inn =0.
\ee
The integral is over the whole spacetime. A particular order of limits is implied here. The $\inn$ and $\out$ states have to be generated first using the LSZ reduction. Hence there is no subtlety in taking derivative inside the matrix element. The final result is going to be proportional to a momentum conservation delta function, imposing
\be
q+\sum_k \eta_k p_k =0
\ee
where $k$ runs over all external states whose on-shell momenta are $\{p_k\}$; $\eta_k = 1$ for outgoing states and $-1$ for ingoing ones; $q$ is an on-shell soft momentum, namely $|q^0|=|\q|\ll E_k\equiv p_k^0$ for all $k$. In what follows we drop this overall delta function and assume that the difference between the total in and total out momenta is $q$. Using the explicit form of the current we get
\be\label{dK1}
\int d^4x\ e^{-iq\cdot x} \left[\d_i\alpha \out \Box A_i(x) - \d_i\d_jA_j(x) \inn
+iq_\mu \alpha \out J^\mu(x) \inn  \right]=0.
\ee
The first term, which is linear in the Maxwell field, is called a ``soft term''. We next use the classification \eqref{vep} for the adiabatic modes. It is seen that the part proportional to $\d_i\d_jA_j$ can be ignored because of the adiabaticity and transversality conditions \eqref{trace} and \eqref{trans}. Then we take $\d_i \alpha$ outside of the integral:
\be\label{d/dq}
\int d^4x\ e^{-iq\cdot x} \d_i\alpha \out \Box A_i(x)\inn
=\sum_{n=0}^{\infty} \frac{(i)^n}{n!}\bvep_{i_0\cdots i_n} \frac{\d^n}{\d \q_{i_1}\cdots\d \q_{i_n}}
\int d^4x\  e^{-iq\cdot x} \out \Box A_{i_0}(x) \inn.
\ee
The integral on the r.h.s. is the standard LSZ formula for creating in- or out-states. Choosing the momentum $q^0>0$, we get the amplitude for the emission of an outgoing soft photon:
\be
\int d^4x\  e^{-iq\cdot x} \out \Box A_{i}(x) \inn = i \langle {\rm out},A_i(\q)\inn.
\ee

\begin{figure}[t]
\centering
\includegraphics[scale = 1]{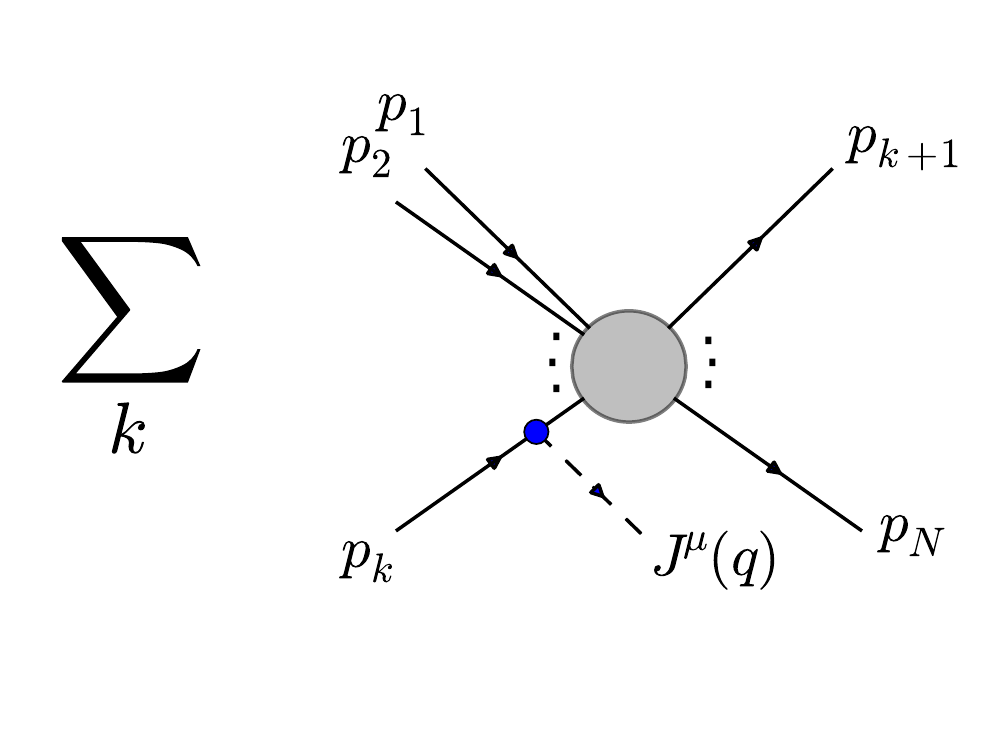} 
\caption{\small{The insertion of the quadratic piece of the electric current in the external lines.}}\label{fig:soft}
\end{figure}

Next consider the second term in \eqref{dK1} which is the contribution from the hard modes. The crucial step here is that $J^\mu$ acts as an interaction Hamiltonian with quadratic and higher order vertices. That is why integration by parts in this term is allowed: asymptotically we turn off the interactions by the $i\ep$ prescription and hence the nonlinear terms in the current vanish (see appendix \ref{app:LSZ} for more details). As in \eqref{d/dq} the $\x$-dependence in $\alpha$ can be taken outside of the integral and replaced by a differential operator. So we are left with 
\be
iq_\mu \sum_{n=0}^{\infty} \frac{(i)^{n+1}}{(n+1)!}\bvep_{i_0\cdots i_n} \frac{\d^{n+1}}{\d \q_{i_0}\cdots\d \q_{i_n}}
\int d^4x e^{-iq\cdot x} \out J^\mu(x) \inn.
\ee
The dominant contribution of this term in the $q\to 0$ limit comes from inserting the quadratic piece of $J^\mu$ in the on-shell external lines as in figure \ref{fig:soft}. This results in a nearly on-shell internal line whose propagator diverges in the $q\to 0$ limit:
\be
\frac{-i}{(p_k+\eta_k q)^2 + m_k^2} = \frac{-i\eta_k}{2q\cdot p_k} + \O(1).
\ee
Other insertions of the current are regular in this limit, therefore
\be\label{outJin}
e^{i\sum\eta_k p_k \cdot x} \out J^\mu(x) \inn = \sum_k \langle k| J^\mu(x)|k\rangle 
\frac{-i\eta_k}{2q\cdot p_k} \langle {\rm out}\inn +\O(1).
\ee
Here we used the fact that in- and out-states are eigenstates of the quadratic piece of $J^\mu(x)$ with eigenvalues
\be
\langle k| J^\mu(x)|k\rangle = 2 Q_k p_k^\mu.
\ee
Putting everything together we finally get the desired soft theorems
\be\label{softs}
\sum_{n=0}^{\infty} \frac{1}{n!}\bvep_{i_0\cdots i_n}\frac{\d^n}{\d \q_{i_1}\cdots\d \q_{i_n}}
\langle {\rm out}, A_{i_0}(q)\inn 
=\sum_{n=0}^{\infty} (-1)^{n} \sum_k \frac{ \eta_k Q_k \bvep_{i_0\cdots i_n}\p_k^{i_0}\cdots \p_k^{i_n}}{(q\cdot p_k)^{n+1}}
\langle {\rm out}\inn,
\ee
In particular for $n=0$ we get the familiar Weinberg formula
\be\label{Wein}
\bvep_i \langle {\rm out}, A_{i}(q)\inn = 
\sum_k \frac{ \eta_k Q_k \bvep_i \p_k^i}{q\cdot p_k}\langle {\rm out}\inn .
\ee
The transversality condition $\q_i  \bvep^i =0$ implies that for any soft momentum, say $q=\omega(1,0,0,1)$, there are two independent adiabatic modes at this order $\bvep_{\pm} =(\hat \x\pm i \hat {\bsb y})/\sqrt{2}$.\footnote{The Weinberg theorem is usually written in a covariant form using a polarization four-vector $\vep_\mu$. It is not possible to choose polarization vectors for all $\q$ to be universally transverse ($q_\mu \vep^\mu=0$) and at the same time purely space-like ($\vep^0=0$). However, it is possible to decompose them as a sum of one such vector $\tilde \vep^\mu$ and a longitudinal one: $\vep^\mu = \tilde\vep^\mu + (\vep^0 /q^0)q^\mu $. The longitudinal piece collapses the r.h.s. of the Weinberg theorem to $\sum_k\eta_k Q_k$ which vanishes by total charge conservation.}

\subsubsection{Redundancy of higher order soft theorems}

From \eqref{softs} there appears to be infinitely many conditions for different configurations of the soft photon. However, they all follow from the Weinberg theorem.\footnote{So there doesn't seem to be any direct relation between the higher order identities in \eqref{softs} and the Low's sub-leading soft theorem \cite{Low,Strominger_subleading}.} Expanding the free out-field in terms of creation and annihilation operators 
\be\label{modes}
A^{\rm out}_\mu(\x) = \int \frac{d^3\q}{(2\pi)^3 2\omega_q} 
\left[\sum_s \vep_\mu^*(s,\q)a_{\rm out}(s,\q)  e^{iq.x}+\text{c.c.}\right]
\ee
we can rewrite \eqref{Wein} as
\be\label{Wein2}
\out a_{\rm out}(s,\q)\inn = 
\sum_k \frac{ \eta_k Q_k \vep_\mu(s,\q)\cdot p^\mu_k}{q\cdot p_k}\langle {\rm out}\inn .
\ee
Substituting the same decomposition in the l.h.s. of other soft theorems in \eqref{softs} and using \eqref{Wein2} gives
\be
\begin{split}
\frac{1}{n!}\bvep_{i_0\cdots i_n} &\frac{\d^n}{\d \q_{i_1}\cdots\d \q_{i_n}}
\sum_s \vep^*_{i_0}(s,\q)\out a_{\rm out}(s,\q)\inn
\\[10pt]
&=\frac{1}{n!} \bvep_{i_0\cdots i_n}\frac{\d^n}{\d \q_{i_1}\cdots\d \q_{i_n}} 
\sum_s \vep^*_{i_0}(s,\q)
\sum_k \frac{ \eta_k Q_k \vep_\mu(s,\q) p^\mu_k}{q\cdot p_k}\langle {\rm out}\inn. 
\end{split}
\ee
Dropping the longitudinal part of $\vep_\mu(s,\q)$ to make it space-like and then using the completeness relation
\be\label{complete}
\sum_s \vep^*_{i}(s,\q)\vep_{j}(s,\q)=\delta_{ij}-\hat \q_i \hat\q_j
\ee
and the fact that 
\be\label{hatq}
\bvep_{i_0\cdots i_n} \frac{\d^m}{\d q_{i_1}\cdots\d q_{i_m}} \hat \q_{i_0} =0,\qquad \text{for all $m$}
\ee 
which follows from \eqref{trace} and \eqref{trans}, we obtain the r.h.s of \eqref{softs}.

\subsection{Adiabatic modes and local correlation functions}

The above symmetries imply an infinite number of constraints on local correlation functions, in contrast to the $\S$-matrix elements where all Ward identities degenerate into the Weinberg theorem. These relations are closer analogs of cosmological consistency conditions. In this section we will derive them using two slightly different methods. 

\subsubsection{Derivation 1: Current conservation}

We start from the requirement of current conservation in a time-ordered correlation function with operator insertions at finite $\{x_k\}$. Unlike \eqref{dK2} there  will be contact terms:
\be\label{dK}
\int d^4x e^{-iq\cdot x} \d_\mu\expect{\hat T\{K^\mu(x) \phi(x_1)\cdots\phi(x_N)\}}
=i\int d^4x e^{-iq\cdot x}\sum_k \delta^4(x-x_k)
\expect{\hat T\{\phi(x_1)\cdots \delta\phi(x_k)\cdots\}}
\ee
where $q$ is an on-shell soft momentum and the variation $\delta\phi$ is defined as
\be
[K^0(t,\x),\phi(t,\x')]= i\delta\phi(t,\x)\delta^3(\x-\x').
\ee
Since $K^0=-\d_i \alpha \d_0 A_i +\alpha J^0$ the transformation of $\phi$ field is the same as a $U(1)$ gauge transformation with parameter $\alpha$. This is what we expect since $\int d^3\x K^0$ is the generator of large gauge transformation with $\delta\phi = iQ\alpha \phi$. 

The l.h.s. contains an LSZ pole and leads to a similar expression as the l.h.s. of \eqref{softs} with an outgoing (ingoing) soft photon if $q^0>0$ ($q^0<0$). To insure that the operator insertions and time-ordering do not affect this result we will give a detailed derivation in appendix \ref{app:LSZ}. Taking the $q\to 0$ limit in the r.h.s. of \eqref{dK} and transforming the spatial coordinates to Fourier space gives
\be
\label{ward}
\begin{split}
\lim_{q\to 0}\sum_n \frac{(i)^{n+1}}{n!}&\bvep_{i_0\cdots i_n}\frac{\d^{n}}{\d \q_i^{i_1}\cdots \d \q^{i_n}}
\langle A^{\rm out}_{i_0}(\q)|\hat T\{\phi(t_1,\p_1)\cdots \phi(t_N,\p_N)\}|0\rangle\\[10pt]
&= \sum_n \frac{i^{n+1}}{(n+1)!}
\bvep_{i_0\cdots i_n} \sum_k Q_k \frac{\d^{n+1}}{\d \p_k^{i_0}\cdots \d \p_k^{i_n}}
\expect{\hat T\{\phi(t_1,\p_1)\cdots \phi(t_N,\p_N)\}}.
\end{split}
\ee

\subsubsection{Derivation 2: Operator product expansion}

Now let us start from the l.h.s. of \eqref{ward} and derive the r.h.s. using the Operator Product Expansion (OPE). For small $\q$ the l.h.s. is related through LSZ formula to the Fourier transform with respect to $x$ of the time-ordered product of $A_i(x)$ and a collection of operators $\{\phi(x_k)\}$. Since the insertions $\{x_k\}$ are very close together compared to $1/|\q|$, we can use OPE to rewrite this product in terms of a sum of operators at say the mid-point $\bar x = \sum_k x_k/N$, which we take to be the origin $\bar x =0$. Because of the existing one-photon state the relevant terms in the OPE are 
\be\label{ope}
\hat T\{\phi(x_1)\cdots \phi(x_N)\} \sim \sum_{n=0}^{\infty}
c_{i\mu_1\cdots \mu_n}\d^{\mu_1}\cdots\d^{\mu_n}A_i(0)+\cdots
\ee
where the non-covariant form is because we work in the $A_0=0$ gauge. Taking the matrix element of \eqref{ope} between in vacuum and the out-state $\langle A^{\rm out}_i(\q)|$ yields
\be
\langle A^{\rm out}_i(\q)|\hat T\{\phi(x_1)\cdots \phi(x_N)\}|0\rangle\sim 
(-i)^n\sum_n c_{j\mu_1\cdots\mu_n} q^{\mu_1}\cdots q^{\mu_n} (\delta_{ij}-\hat \q_i\hat\q_j).
\ee
where we used the mode expansion \eqref{modes} and completeness relation \eqref{complete}. Applying the derivative operator on the l.h.s. of \eqref{ward}, and using \eqref{hatq} we get 
\be\label{project}
i\sum_n c_{i_0\cdots i_n} \bvep_{i_0\cdots i_n}.
\ee
Next we derive this particular projection of the OPE coefficients. Generally, the OPE coefficients in \eqref{ope} are the linear response of the correlation function of $\{\phi(x_k)\}$ to a background gauge field $A_i$:\footnote{They can in principle be calculated using the background field method in the path integral over a region including all insertion points $\{x_k\}$.} 
\be\label{response}
\expect{\hat T\{\phi(x_1)\cdots \phi(x_N)\}}_{A_i}-\expect{\hat T\{\phi(x_1)\cdots \phi(x_N)\}}_{A_i =0}.
\ee
Higher values of $n$ correspond to the response to higher derivatives of the background field and hence are suppressed by additional factors of $q\cdot (x_i -x_j)\ll 1$. Suppose in the vicinity of $\{x_k\}$ the long wavelength Maxwell field is locally removable by a $U(1)$ gauge transformation $A_i(x)= \d_i \alpha(x)$, namely it is an adiabatic mode. Then the linear response is trivially given by the inverse transformation of the charged operators $\delta\phi(x_k)=-iQ_k \alpha(x_k)\phi(x_k)$:
\be
\sum_n c_{i\mu_1\cdots \mu_n}\d^{\mu_1}\cdots\d^{\mu_n}\d_i\alpha(0)
= -i \sum_k Q_k \alpha(x_k)\hat T\{\phi(x_1)\cdots \phi(x_N)\} .
\ee
Plugging in the expansion \eqref{vep} allows us to determine the projection of the OPE coefficients for adiabatic modes. The l.h.s. is $-i$ times the desired projection \eqref{project}. Transforming $\{\x_k\}$ to Fourier space, the r.h.s. becomes $-i$ times the r.h.s. of \eqref{ward}. 

In conclusion, the infinite set of symmetries impose an infinite set of conditions on the OPE coefficients of charged fields fusing into the adiabatic configurations of the $U(1)$ gauge field. It is worth mentioning that the Weinberg theorem cannot be derived by a simple application of the LSZ formula to \eqref{ward}. LSZ requires integrating the time-ordered expectation value over the whole spacetime to create in and out states, but the OPE does not hold in this case. This is also consistent with the fact that the Weinberg formula \eqref{Wein} has a pole in the soft momentum. The external hard modes travel for long enough time to experience the long wavelength modulation of the soft photon. This is outside the regime where the local ward identity holds.

\section{Adiabatic modes in gravity}

Gravitational adiabatic modes in asymptotically flat spacetime can be defined in a similar way as in electrodynamics, again as a simple generalization of \cite{Weinberg_adia,Hinterbichler2}. On Minkowski background the theory is invariant under infinitesimal diffeomorphisms. Define canonically normalized metric fluctuation by writing $g_{\mu\nu} = \eta_{\mu\nu} +\kappa h_{\mu\nu}$, where $\kappa^2 = 32\pi G$. Under a diffeomorphism $\xi^\mu$ it transforms nonlinearly
\be
h_{\mu\nu}\to h_{\mu\nu} + \kappa^{-1}\d_\mu\xi_\nu +\kappa^{-1} \d_\nu\xi_\mu + \xi^\sigma\d_\sigma h_{\mu\nu},
\ee
while fields with spacetime independent VEV transform linearly. To fix local diffeomorphisms we impose the synchronous gauge
\be\label{gauge}
h_{00} = h_{0i} =0.
\ee
Transverse, time-independent large diffeomorphisms $\d_0\xi_i^T = \d_i\xi_i^T=0$ preserve this gauge. \footnote{A larger group of large diffeomorphisms preserve this gauge. However, unless there is a dynamical longitudinal degree of freedom in the metric (as in cosmology) they won't lead to adiabatic modes.} Acting on Minkowski vacuum these large diffeomorphisms generate a class of infinite wavelength solutions. A subclass of them can be continued to physical solutions. To see this let us investigate the linearized Einstein equations:
\be\label{Einstein}
\Box h_{\mu\nu}-\d_\mu\d_\sigma h^\sigma_\nu-\d_\nu\d_\sigma h^\sigma_\mu +\d_\mu\d_\nu h - \eta_{\mu\nu}\Box h
+\eta_{\mu\nu}\d_\sigma\d_\rho h^{\sigma\rho}= \frac{\kappa}{2}T_{\mu\nu}.
\ee
The $\{0i\}$ component acts as a constraint and in the absence of sources reads
\be
\d_0(\d_i h_{kk} - \d_k h_{ik}) =0.
\ee
Any $h_{ij} = \d_i\xi^T_j + \d_j \xi^T_i$ with $\d_0\xi^T_i =0$ satisfies this equation. However to guarantee continuity to finite frequency when $\d_0\neq 0$ we impose the stronger requirement that $(\d_i h_{kk} - \d_k h_{ik}) =0$. This forces
\be\label{adia}
\nabla^2 \xi_i^T =0.
\ee
We can again organize large diffeomorphisms in a Taylor series
\be\label{xi}
\xi^T_{i} =  \sum_{n=0}^\infty \frac{1}{(n+1)!} \bvep_{i i_0 i_1\cdots i_n} \x^{i_0}\cdots \x^{i_n},
\ee
where the matrices $\bvep_{i i_0\cdots i_n}$ are maximally symmetric in their last $n+1$ indices. The adiabaticity condition \eqref{adia} implies the trace condition
\be\label{adia2}
\bvep_{ijj j_1\cdots j_n}=0.
\ee
Moreover, after continuation to finite momentum, requiring a soft mode of momentum $q$ to be transverse imposes an additional constraint
\be\label{trans2}
\q_i \bvep_{(i j)\cdots}\equiv \q_i (\bvep_{i j\cdots}+\bvep_{ji\cdots})=  0.
\ee
Next we derive the Noether currents associated to these symmetries and show how they lead to the soft graviton theorem.

\subsection{Conserved currents and soft graviton theorem}

Having found a symmetry of the gauge fixed action one can derive the Noether current by varying the action as in appendix \ref{app:Noether}, or directly from the equations of motion. We use the equations of motion, which after linearizing in $h_{\mu\nu}$ can be written as
\be\label{gam}
\d^\alpha H_{\alpha \mu\nu} = \frac{\kappa}{2} T_{\mu\nu},\quad \text{with}\quad 
H_{\alpha\mu\nu} = \d_\alpha h_{\mu\nu}+ \eta_{\nu\mu}\d^\beta h_{\alpha\beta} + \eta_{\nu\alpha}\d_\mu h^{\beta}_{\beta}
-\{\alpha\leftrightarrow \mu\}.
\ee
Linearization in $h_{\mu\nu}$ is not valid in the presence of hard gravitons. However, since the final soft theorem only depends on asymptotic hard states the energy-momentum of hard gravitons can be included in $T_{\mu\nu}$. From this we can deduce the Noether current by requiring that its divergence results in a projection of \eqref{gam} onto the adiabatic mode:
\be\label{K}
K^\mu = \d_i \xi^T_j H^{\mu i j} - \frac{\kappa}{2} \xi^T_i \ T^\mu_i .
\ee
The conservation of this current in the presence of in- and out-states leads to the soft graviton theorems. We start with
\be\label{dKgr}
0 = \int d^4x e^{-iq\cdot x} \d_\mu \out K^\mu \inn 
= \int d^4x e^{-iq\cdot x} \left[\d_{i}\xi^T_{j} \out \Box h_{ij} \inn
-i\frac{\kappa}{2}q_\mu \xi^T_i\out T_i^\mu \inn \right],
\ee
where we used
\be
\d_i\xi^T_j \d_\mu H^{\mu ij} = \frac{1}{2}\d_{(i}\xi^T_{j)} \d_\mu H^{\mu ij},
\ee
the transversality of $\xi^T_i$, \eqref{adia} and \eqref{trans2} to simplify the soft term. The first term on the r.h.s. creates a soft outgoing graviton, assuming that $q^\mu$ is a positive frequency null momentum. $\d_{i}\xi^T_{j}$ projects it onto adiabatic configurations. 

We then use the fact that the stress-energy tensor acts as a quadratic interaction vertex and hence leads to a nearly on-shell propagator with a $1/q$ singularity when inserted in external on-shell lines. Therefore
\be
e^{i\sum_k \eta_k p_k\cdot x} \out T^{\mu\nu}(x) \inn = \sum_k \langle k| T^{\mu\nu}(x)|k\rangle 
\frac{-i\eta_k}{2q\cdot p_k} \langle {\rm out}\inn +\O(1).
\ee
The in- and out-states are eigenstates of the quadratic piece of $T^{\mu\nu}(x)$ with eigenvalues
\be\label{kTk}
\langle k| T^{\mu\nu}(x)|k\rangle = -2 p_k^\mu p_k^\nu.
\ee
If terms of $\O(h_{\mu\nu}^2)$ were kept in the current $K^\mu$, they would correspond to the energy-momentum tensor of free gravitons. Hence the same equation \eqref{kTk} also applies to ingoing and outgoing hard gravitons. Substituting the expansion \eqref{xi}, we find
\be\label{gsofts}
\sum_{n=0}^{\infty} \frac{1}{n!}\bvep_{ijj_1\cdots j_n} \frac{\d^n}{\d \q_{j_1}\cdots\d \q_{j_n}}
\langle {\rm out}, h_{ij}(q)\inn 
=\frac{\kappa}{2} \sum_{n=0}^{\infty} (-1)^{n} \sum_k \frac{ \eta_k \bvep_{i j\cdots j_n}\p_k^{i}\p_k^{j}\cdots \p_k^{j_n}}
{(q\cdot p_k)^{n+1}}
\langle {\rm out}\inn,
\ee
where $\eta_k = 1$ for outgoing particles and $-1$ for ingoing ones. In particular for $n=0$ we get the Weinberg soft theorem
\be
\bvep_{ij} \langle {\rm out}, h^t_{ij}(q)\inn = 
\frac{\kappa}{2}\sum_k \frac{ \eta_k \bvep_{ij} \p_k^i \p_k^j}{q\cdot p_k}\langle {\rm out}\inn ,
\ee
As in electrodynamics higher order identities in \eqref{gsofts} can be derived starting from this relation and a mode expansion of $h^t_{ij}$ in terms of definite helicity states. Hence they are not independent. 

\section{BMS, continued to a finite radius}\label{sec:BMS}

The goal of this section is to make connection between the above derivation of soft theorems and the recent derivations based on the asymptotic symmetry groups of electrodynamics \cite{Strominger_photon} and gravity (BMS) \cite{Strominger_graviton}. For simplicity, we only consider soft photon Ward identities, although we expect our arguments to apply to the gravitational case with minor modifications. Using the conservation of the currents associated to the adiabatic modes, we will derive conserved charges defined on the world-volume of a big sphere of radius $R$ that surrounds the scattering event. There is one conserved charge associated to every angle $\hat \n$. We will show how in the $R\to \infty$ one can obtain the asymptotic charges defined in \cite{Strominger_photon} for massless QED, \cite{Strominger_QED} for massive QED, and their higher dimensional generalization \cite{Strominger_even}. In particular, we give a proof of the antipodal matching condition. Finally, we will comment on the proposal of \cite{Susskind}. 

We use spherical coordinates to describe a scattering event taking place near the origin, 
\be
ds^2 = -dt^2 +d\rho^2 + \rho^2 \gamma_{ab}dz^a dz^b,
\ee
where $\gamma_{ab}$ is the metric of the unit $(d-2)$-sphere. Covariant derivative with respect to $\gamma_{ab}$ is denoted by $D_a$. To discuss radiation we often switch to retarded (advanced) variables $u=t-\rho$ ($v=t+\rho$), $r=\rho$. However, index $0$ always refers to $t$.\footnote{Since $d=4$ is in some respects special we keep $d$ arbitrary to emphasize the generality of our arguments. We often use the shorthand 
\be
d\hat\r \equiv d^{d-2}z \sqrt{\gamma}.
\ee } 

As in section \ref{sec:photon} we fix the Maxwell field by choosing the temporal gauge. This implies that $A_0=A_u=A_v=0$ and $A_r = A_\rho$. This gauge choice makes comparison with the works on asymptotic symmetry groups easier and is the same choice as in \cite{Susskind}. Since our goal is to work at a finite distance the choice $A_r=A_u|_{\scri^+}=0$ used in \cite{Strominger_photon} seems inconvenient. 

The adiabatic modes were identified in section \ref{sec:photon} as time-independent large gauge transformations $\alpha(\r)$ that are harmonic $\nabla^2\alpha(\r) =0$ (note that $\d_0\alpha =0$ implies $\d_r \alpha = \d_\rho \alpha$). The Noether current is
\be
K_\mu = g^{ij}\d_i\alpha F_{\mu j} + \alpha J_\mu.
\ee
The current conservation implies that the charge defined by integrating the current over any closed co-dimension one surface that does not surround an operator insertion must vanish. That is, the following Heisenberg operator is zero
\be
\Q = \int_M d^dx \d_\mu (\sqrt{-g} K^\mu) = \oint_{\d M} d^{d-1}\Sigma\ n^\mu K_\mu =0,
\ee
where $M$ is an arbitrary region in spacetime, $d\Sigma$ is a differential element of its boundary $\d M$, and $n^\mu$ is the normal to $\d M$. The vanishing of the matrix elements of this operator between Heisenberg picture in- and out-states gives us the conservation laws.\footnote{We often use $\Q$ to also denote these matrix elements.} 

We choose $M$ to be as in figure \ref{cyl}: the region enclosed by a cylinder made of the world-volume of a sphere of radius $R$ surrounding the scattering event at $t=\rho=0$, and capped by two space-like surfaces at a very late time $T$ and a very early time $- T$ with $T\gg R$. The surface integral is
\be\label{Qdef}
\begin{split}
\Q=&\left.\int d\hat \r \Big[\int_0^R d\rho \rho^{d-2} \alpha(\rho \hat \r) 
(-\d_0 \nabla^i A_i+J_t)\right|_{-T}^{T}\\[10pt]
&~~~~~~~~~~+R^{d-2}\alpha(R \hat \r) \int_{-T}^{T} dt (-R^{-2} D^aF_{\rho a}  +J_\rho)\Big]=0,
\end{split}
\ee
where in the first term $\nabla$ denotes the covariant derivative with respect to $g_{ij}$, and we used the antisymmetry of $F_{\mu\nu}$ in the second term. It should be noted that using the Maxwell equations the current can be written as a total divergence of an antisymmetric tensor:
\be\label{exact}
K^\mu = \frac{1}{\sqrt{-g}}\d_\nu(\sqrt{-g}\alpha F^{\mu\nu}),
\ee
i.e. it is an exact form: $K= *d*(\alpha F)$ where $d$ is exterior derivative and $*$ is Hodge dual operator. Hence, the integral of $K_\mu$ over a $3d$-surface reduces to a $2d$ integral over its boundaries, from which the vanishing of $\Q$ defined on any closed surface $\d M$ follows. 

Note also that our large gauge transformations are different from those in \cite{Strominger_photon} which would correspond to $r$-independent $\beta(\hat \r)$. The latter leads to an $r$-independent field $A_a = \d_a \beta$, which has the same large $r$ scaling as radiation in $4d$, but it is superdominant at higher dimensions where $A^{\rm rad}_a \propto r^{2-d/2}$. On the other hand, there is no relation between the $r$-dependence of adiabatic modes $A_i = \d_i \alpha$ and soft radiation or the boundary conditions. As discussed in section \ref{sec:photon} adiabatic modes characterize locally unobservable configurations which can be deformed at larger distances. Here there is no coincidence that makes $4d$ different from higher dimensions. 

Once $R$ is fixed our $\alpha(R\hat\r)$ can be chosen to have an arbitrary $\hat\r$-dependence. The condition $\nabla^2 \alpha$ is satisfied by an appropriate choice for the $r$-dependence. Therefore, when integrating over the world-volume of the sphere, we replace $\alpha(R\hat\r)\to \alpha(\hat\r)$, and eventually, we will obtain the same conservation laws for the actual scattering processes as in \cite{Strominger_photon}. This is the subject of next few sections.

\begin{figure}[th!]
\centering
\includegraphics[scale = 0.7]{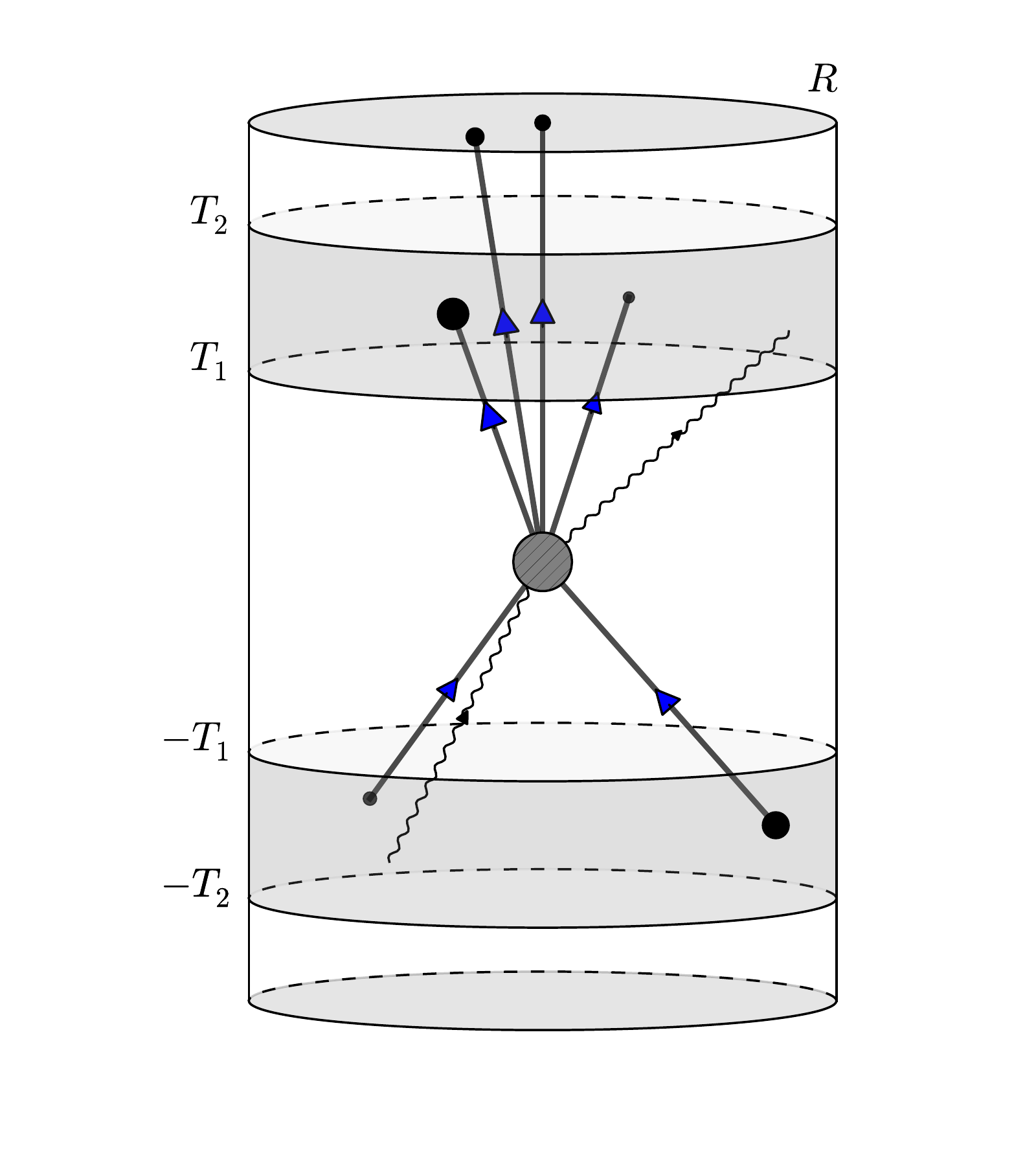} 
\caption{\small{The world-volume of the sphere of radius $R$ on which the charge $\Q$ is defined. Ingoing and outgoing massless particles and radiation enter and exit the enclosed region $M$ respectively through the shaded areas $(-T_2,-T_1)$ and $(T_1,T_2)$. For large enough $R$ massive states enter well before $-T_2$ and exit well after $T_2$.}}\label{cyl}
\end{figure}

\subsection{Massless QED}\label{sec:qed}

Suppose all ingoing and outgoing states are massless. Then by our choice $T\gg R$ the first term in \eqref{Qdef} vanishes since radiation and massless particles enter and exit region $M$ at $t\sim \pm R$ (see figure \ref{cyl}). Moreover, we can use the fact that for massless outgoing particles $J_\rho = -J_t=-J_u$ and for ingoing ones $J_\rho = J_t=J_v$. Splitting the time integral into one over $t>0$ and one over $t<0$ and switching respectively to retarded and advance coordinates, we find
\be\label{Quv}
\begin{split}
\Q\equiv\Q_+ - \Q_-=R^{d-2}\int d\hat \r \alpha(\hat \r)
&\Big[\int_{-R}^{T} du [R^{-2}(\d_u D^aA_a-\d_r D^aA_a + D^2 A_r) -J_u]\\[10pt]
&-\int_{-T}^{R} dv [R^{-2}(\d_v D^aA_a+\d_r D^aA_a - D^2 A_r) -J_v]\Big]=0.
\end{split}
\ee
Modulo the difference in the gauge choice, in the limit $T,R\to \infty$ this expression becomes almost identical to the asymptotic charges defined in \cite{Strominger_photon}. However, there is no antipodal matching: Choosing the transformation parameter $\alpha(\hat \r)$ to peak at a specific direction on the sphere, the charge gets contribution from that same direction all the way from $-T$ till $+T$. Given that the asymptotic conservation laws of \cite{Strominger_photon} are shown to be equivalent to the Weinberg theorem there must be a way to understand how they follow from the above conservation laws. We will argue that antipodal matching in \cite{Strominger_photon} can be derived by including the contribution from the dressing Maxwell fields of the in- and out-states. Note that without the dressing field even a freely moving charge cannot satisfy \eqref{Quv}, since it doesn't radiate but its electric current $J_\mu$ contributes as we see below.

Let us divide the charge $\Q$ into a hard $\Q_H$ piece which is coming from the electric currents and a soft $\Q_S$ piece from the Maxwell field. Using semi-classical expression for the current associated to massless particles
\be
\begin{split}
J_u =\frac{1}{r^{d-2}} \sum_{k\in {\rm out}} Q_k \delta(u-u_k) \delta(\hat \r-\hat \r_k)\\[10pt]
J_v =\frac{1}{r^{d-2}} \sum_{k\in {\rm in}} Q_k \delta(v-v_k) \delta(\hat \r-\hat \r_k)
\end{split}
\ee
we get
\be\label{QHH}
\Q_H =- \sum_{k} \eta_k Q_k \alpha(\hat \r_k)\  \out\innn\rangle,
\ee
where as before $\eta_k=1$ for outgoing states and $-1$ for ingoing ones. 

As for the soft piece, it should itself be divided into a radiative contribution $\Q_S^{\rm rad}$ and a contribution from the dressing fields $\Q_S^{\rm dr}$. The latter is nonzero because any quantum state in Maxwell theory has to satisfy the Gauss constraint, which implies that the in (and similarly out) states with charged particles have to be dressed by a classical Maxwell field $\A_\innn$, not to be confused with the quantum field $A_i$, \cite{Dirac,Bagan}:
\be
|{\rm in}, \A_\innn\rangle = \exp\left[i\int d^{d-1}\x (\A_\innn^i \d_0{ A_i}
+{ \E_\innn}^i   A_i)\right] \inn
\ee
where $\E_\innn =-\d_0 \A_\innn$ and $\nabla\cdot { \E}_\innn =  J^0_{\rm in}$. There is a freedom in the choice of $\A_\innn$ analogous to the freedom in the choice of boundary conditions in classical electrodynamics. Both $\Q_S^{\rm rad}$ and $\Q_S^{\rm dr}$ depend on the choice, but the sum of the two is unambiguous. 

In quantum electrodynamics the default choice for the dressing field is that of free moving in (out) states without any radiation. That's why when discussing scattering processes with emission and absorption of finite frequency photons there is no need to consider the dressing fields. The original derivation of the soft theorems by Weinberg and the derivation given in section \ref{sec:photonwein} followed the same logic by working at a finite frequency (recall the factor of $e^{-iq\cdot x}$), and then taking the soft limit. Here, on the other hand, we are considering photons whose wavelengths are longer than the box size $\omega R\ll 1$ --- in the limit $R\to \infty$ where the asymptotic charges are defined we are exclusively talking about the zero-mode. Hence the contribution from the dressing fields has to be taken into account.

We do not need a detailed treatment of the dressing field. All that we need is to determine $\Q^{\rm dr}_S$ which can be done by requiring $\Q =0$ for trivial processes. Consider a forward scattering $\inn \to |{\rm out = in}\rangle$. The vanishing of $\Q$ implies
\be\label{Qin}
\langle {\rm out}={\rm in}, \A_{\rm in}| \Q |{\rm in},  \A_{\rm in}\rangle 
= \Q^{\rm dr}_S(\A_{\rm in}) + \sum_{\rm in} Q_k [\alpha(\hat \r_k)-\alpha(-\hat \r_k)]\langle{\rm in}\inn = 0.
\ee
We defined $\Q^{\rm dr}_S(\A_\innn)$ as the dressing contribution and used \eqref{QHH} and the fact that in this process every ingoing charge enters at $\hat \r_k$ and exits at $-\hat \r_k$.\footnote{By unitarity $\langle{\rm in}\inn \neq 1$ as soon as there is a nonzero scattering cross section.} Using the same prescription for the out-states gives a similar expression for $\Q_S^{\rm dr}(\A_{\rm out})$ in terms of the charges. In appendix \ref{app:dress} we show explicitly how the field of a freely moving massless charge cancels the contribution from its electric current to $\Q$.

When the ingoing particles evolve forward and scatter into the outgoing ones, the Maxwell field changes from $\A_\innn$ due to outgoing electromagnetic radiation. Similarly, as the out-state evolves backwards into the in-state, its field $\A_{\rm out}$ changes since there will be ingoing radiation. We are interested in separating the radiative contribution. Thus $\Q_S^{\rm dr}$ is what we get if $\A_\innn$ freely evolves until late times and $\A_{\rm out}$ freely evolves to early times. Since the real part of the matrix element of $A_i$ between two coherent states with $\A_{\innn}$ and $\A_{\rm out}$ is the average, it follows for a generic matrix element that
\be\label{weak}
\begin{split}
\langle {\rm out}, \A_{\rm out}| \Q |{\rm in},  \A_{\rm in}\rangle 
&= \Q_S^{\rm rad}+\frac{1}{2}\left[\frac{\Q^{\rm dr}_S(\A_{\rm in})}{\langle {\rm in}| \innn\rangle} 
+\frac{\Q^{\rm dr}_S(\A_{\rm out})}{\out {\rm out}\rangle}\right]\out\innn\rangle +\Q_H\\[10pt]
&=\Q_S^{\rm rad}+\frac{1}{2}\sum_k \eta_k Q_k [\alpha(\hat \r_k)-\alpha(-\hat \r_k)]\out\innn\rangle = 0.
\end{split}
\ee
where $\Q_S^{\rm rad}$ is the radiative contribution to the soft terms in \eqref{Quv}. This is a Ward identity that relates the amplitudes of scattering processes with and without soft photon emission. Using the same techniques used in \cite{Strominger_photon}, \eqref{weak} can be derived from Weinberg soft theorem (appendix \ref{app:soft}). This is true for arbitrary $\alpha(\hat \r)$ and hence there is one conservation law for every direction. However, they are weaker than the Weinberg soft theorem, and also weaker than the identities derived in \cite{Strominger_photon} with antipodal matching. To see this fix $\alpha(\hat\r) = \delta(\hat\r-\hat \n)$ and note that 

(a) The vanishing of $\Q(\n)$ relates $\Q_{S+}^{\rm rad}(\hat \n)-\Q_{S-}^{\rm rad}(\hat \n)$, i.e. the contributions from ingoing and outgoing radiation at $\hat \n$, to the ingoing and outgoing electric flux at both directions $\hat \n $ and $-\hat \n$. On the other hand, Weinberg theorem can be used to derive $\Q_{S+}^{\rm rad}(\hat \n)$ in terms of ingoing electric flux just at $-\hat \n$ and outgoing electric flux just at $\hat \n$. 

(b) Secondly, Weinberg theorem gives the emission/absorption amplitudes for each photon polarization in every direction, while the above relations depend on a combination of two polarizations in a given direction. 

The second weakness is in common with the identities found in \cite{Strominger_photon}. However, it is not a real weakness. An interesting finding of \cite{Strominger_photon} is that at the leading order in $\omega\to 0$ limit one linear combination of the two photon polarizations decouple. That is to say, the zero-mode of the asymptotic Maxwell field $A_a$ has a single degree of freedom instead of two naively expected in four dimensions. It was shown in \cite{Strominger_photon} that the Weinberg formula agrees with this expectation and the amplitude for the two polarizations are linearly dependent. We will next derive the stronger conservation laws to resolve the shortcoming (a). We will also see from a different perspective why soft electromagnetic radiation has only one degree of freedom. 
\subsubsection{Antipodal matching}\label{sec:anti}

Seven technical steps have to be made to derive a version of the asymptotic conservation laws which are as strong as the Weinberg theorem.

{\em 1) Retarded boundary condition:} For any fixed in- and out-states let us choose the (unconventional) retarded dressing field. Namely, the in-state is dressed with the same non-radiative field $\A_\innn$ as before, but the out-state is dressed with the field of the specified in-state plus the a priori unknown soft radiation field that is produced in the scattering $\inn \to |{\rm out}\rangle$:
\be
|{\rm out}, \A_{\rm ret}\rangle = (1+F)|{\rm out}, \A_{\rm in}\rangle
\ee
where the operator $F$ is defined as
\be
F = 
\int_{\omega R<1}\frac{d^{d-1}\q}{(2\pi)^{d-1}2\omega}\sum_s f(s,\q) a^\dagger(s,\q)
\ee
$\omega = |\q|$, and the emission coefficient is defined as
\be\label{alpha}
f(s,\q) = \frac{\out a_{\rm out}(s,\q) \inn}{\out \innn\rangle},
\ee
with no radiation in $\out$ and $\inn$. Similarly we define the absorption coefficient as
\be\label{beta}
g(s,\q) = \frac{\out a_{\rm in}^\dagger(s,\q) \inn}{\out \innn\rangle}.
\ee
We distinguish the matrix elements $\langle{\rm out},\A_{\rm ret}|\Q|\innn,\A_\innn\rangle$ by a superscript $\Q^{\rm ret}$. We will argue that the contribution of incoming soft radiation to $\Q^{\rm ret}_S$, corresponding to incoming soft photons through the interval $(-T_2,-T_1)$ in figure \ref{cyl}, vanishes with this choice, because
\be\label{retarded}
\lim_{\omega\to 0}\langle {\rm out}|(1+F^\dagger)a_{\innn}^\dagger(s,\q)|{\rm in}\rangle = 0.
\ee

{\em 2) Monopole radiation is zero:} Consider the interaction Hamiltonian $H_I(t)=\int d^{d-1}\x A_i J^i$ in $A_0 =0$ gauge. We can use the conservation of the current to write
\be
H_I =\int d^{d-1} \x A_i [\d_j(\x^i J^j) +\x^i \d_0 J^0]. 
\ee
Let us now do a mode expansion of $A_i$ in terms of creation and annihilation operators and take the long wavelength limit.\footnote{We will be working in the interaction picture and not Heisenberg picture until point 3 below.} At leading order 
\be\label{HI}
\lim_{\q \to 0} H_I(t) =  \sum_s [\vep_i^*(s,\q)a(s,\q)+{\rm c.c.}]\int d^{d-1}\x[\d_j(\x^i J^j) +\x^i \d_0 J^0]
+\cdots,
\ee
which is a total derivative. Namely, there is no monopole radiation.

{\em 3) Dipole approximation:} Charges couple to a long wavelength electromagnetic wave via its electric and magnetic field, i.e. the next-to-leading term in \eqref{HI}. This is a coupling of the form
\be
H_{I,\rm dipole} \propto \sum_s [a(s,\q) - a^\dagger (-s,\q)]\vep^*_\mu(s,\q)
\ee
where we made the choice $\vep_\mu(s,\q) = \vep_\mu^*(-s,\q)$. Therefore, $a(s,\q) - a^\dagger (-s,\q)$ commutes with the $\S$-matrix. Dropping $a \inn$ and $\out a^\dagger$ which correspond to disconnected processes, we get
\be\label{commute}
\lim_{\q \to 0} a(s,\q)\ \S =  - \S \ a^\dagger (-s,\q).
\ee
This is almost the desired result \eqref{retarded} written in the interaction picture. Before completing the proof we make another observation. 

{\em 4) Tangential magnetic field decouples:} For large $R$ the charges all move radially for the most part. Therefore they decouple from the tangential component of the magnetic field. Thus in the absence of initial magnetic field which would have evolved trivially, we can set\footnote{Note that if we dualize this tangential $F_{ab}$ to get a magnetic field pseudo-tensor, in $4d$ we get a vector $B_i = \ep_{ijk}F_{jk}$ which corresponds to a radial magnetic field vector $B_r$.}
\be
F_{ab}=0.
\ee
Therefore $A_a =\d_a \varphi$. This is not a pure gauge because $\varphi$ can depend on $t$ and $\rho$, while $A_0 =0$. At large distances the transverse polarizations of electromagnetic radiation lie within the sphere, so this implies that the radiative Maxwell field consists of a single degree of freedom in the soft limit.

{\em 5) Analyticity:} The emission or absorption of soft quanta do not cause internal lines to go from being on-shell to being off-shell or vice versa. Hence the amplitudes $\out a_{\rm out}(s,\q) \inn$ and $\out a^\dagger_{\rm in}(s,\q) \inn$ have the same analytic structure as $\out \innn\rangle$. The ratios \eqref{alpha} and \eqref{beta} are complex only because of the polarization vectors. Using \eqref{commute} and $\vep_\mu(s,\q) = \vep_\mu^*(-s,\q)$ then implies
\be\label{alphabeta}
\lim_{\omega\to0} f(s,\q)=-g^*(s,\q).
\ee
This proves \eqref{retarded}. It follows that $\Q^{\rm ret}$ receives radiative contribution only in the interval $T_1<t<T_2$ in figure \ref{cyl}.

{\em 6) Antipodal incoming radiation:} The Maxwell field in $\Q_{S+} \propto \int_{T_1}^{T_2} dt D^a F_{\rho a}$ can annihilate radiation emitted from the scattering event, and since now there is also radiation in the dressing field of the out-state, it can create photons
\be
\frac{\Q_{S+}^{\rm ret}}{\langle{\rm out}\inn} \supset f^*(s,\q)\langle 0| a(s,\q) \ \int d\hat\r \alpha(\hat \r)
\int_{T_1}^{T_2} dt  D^aF_{\rho a}|0\rangle\neq 0.
\ee
For radiation that goes through the origin we can write this, using \eqref{alphabeta}, as
\be
 g(s,\q)\langle 0| a(s,\q) \ \int d\hat \r \alpha(-\hat\r) 
\int_{-T_2}^{-T_1} dt  D^aF_{\rho a}|0\rangle \subset -\frac{\xoverline\Q_{S-}^{\rm rad}}{\langle{\rm out}\inn}
\ee
where we used the fact the normal to the sphere changes sign at the antipodal point $\hat \rho(\hat\r)=-\hat \rho(-\hat\r)$. $\xoverline \Q_{S-}^{\rm rad}$ is calculated with the conventional boundary condition but at the antipodal point.

{\em 7) Antipodal incoming electric flux:} Finally, we need to add the contribution from the non-radiative dressing field of the in-state $\A_\innn$ which is common between in- and out-states and hence is the same as $\langle{\rm out}\inn/\langle {\rm in}\inn$ times $\Q_S^{\rm dr}(\A_\innn)$ in \eqref{Qin}. Combined together we obtain
\be\label{strong}
\Q^{\rm rad}_{S+} -\xoverline\Q^{\rm rad}_{S -} = \sum_k \eta_k Q_k \alpha(\eta_k\hat \r_k)\langle{\rm out}\inn,
\ee
with the l.h.s. evaluated with the default boundary condition. This is the desired result.

\subsection{Massive QED}\label{sec:massive}

Now consider adding massive scattering particles. As before we are interested in writing the vanishing of $\Q$ as a relation between the electric charges and momenta of the asymptotic states and the soft radiation. 

Suppose the scattering event has a duration $\tau$. We choose $R\gg \tau$. Then most of the ingoing radiation (as well as massless hard states) enter the sphere in the interval $(-T_2,-T_1)$ with $T_1,T_2 \sim R$, $T_2 -T_1 \gg \tau$. Similarly, outgoing radiation leaves the sphere between $T_1$ and $T_2$. 

On the other hand the massive particles keep contributing to $\Q$ at earlier and later times, both via their field and also via their contribution to the electric current --- they eventually have to leave our closed hyper-surface. We use the exactness of the Noether current \eqref{exact} to reduce this part to two surface terms on the $(d-2)$-spheres at $T_2$ and $-T_2$. So the vanishing of $\Q$ implies
\be
\Q(-T_2,T_2) = R^{d-2}\int d\hat\r \alpha(\hat \r)[E_r(T_2,R,z^a)-E_r(-T_2,R,z^a)].
\ee
The electric field of a freely moving charge of velocity $\bsb \beta$ which goes through $\r =0$ at $t=0$ is given by Li\'enard-Wiechert
\be\label{E}
\bsb E=\frac{\gamma Q (\r - \bsb \beta t)}{[\gamma^2(t-\bsb \beta\cdot \r)^2 -t^2 +r^2]^{(d-1)/2}},\qquad
\gamma =\frac{1}{\sqrt{1-\beta^2}}.
\ee
Taking the limit $R\gg (T_2-R)$ in \eqref{E} -- which is the analog of taking the limit $r\to \infty$ and then $u\to \infty$ for outgoing particles or $v\to -\infty$ for ingoing particles in \cite{Strominger_QED} -- we get
\be\label{E2}
-E_r(T_2,R,z^a)+E_r(-T_2,R,z^a) = \frac{1}{(R)^{d-2}\Omega_{d-2}}\sum_k
\frac{\eta_k Q_k}{\gamma_k^{d-2}(1-\eta_k\hat\r\cdot\bsb \beta_k)^{d-2}}
\ee
where the sum runs on the massive particles and $\Omega_{d-2}$ is the area of unit $(d-2)$-sphere. 

This differs from \cite{Strominger_QED} in that the massive in-states are regularly and not antipodally matched to the out-states. However, as before once the contribution from dressing fields to $\Q(-T_2,T_2)$ is included the same result will follow. 

We divide $\Q(-T_2,T_2)$ into the hard and the soft parts. The only novelty compared to the analysis of the previous section is to include dressing fields of massive charges. By considering a trivial process in which a massive in-state freely propagates through the sphere, we find
\be\label{dress}
Q_S^{\rm dr}(-T_2,T_2) = R^{d-2}\int d\hat\r \alpha(\hat\r)[\E_r(T_2,R,z^a)-\E_r(-T_2,R,z^a)].
\ee
To obtain the conservation laws for any fixed in- state take the dressing field of the out-state to be the non-radiative $\A_{\rm in}$ plus the unknown retarded radiation. This eliminates contributions to $\Q^{\rm rad}_{S-}$ but changes $\Q^{\rm rad}_{S+} \to \Q^{\rm rad}_{S+}-\xoverline\Q^{\rm rad}_{S-}$. Moreover, the contribution from $\A_{\rm in}$ is given by \eqref{dress}. Combined together we get
\be\label{Qmass}
\Q^{\rm rad}_{S+} -\xoverline \Q^{\rm rad}_{S -} = \sum_k \eta_k Q_k \alpha(\eta_k\hat \r_k)
+\int d\hat\r \alpha(\hat\r)\sum_k
\frac{\eta_k Q_k}{\Omega_{d-2}\gamma_k^{d-2}(1-\hat\r\cdot\bsb \beta_k)^{d-2}}.
\ee
In summary, we have shown that once the dressing field of in- and out-states are included the asymptotic Ward identities proposed in \cite{Strominger_photon,Strominger_QED,Strominger_even} can be obtained from the conservation of a Noether current associated to adiabatic modes. After deriving the antipodal matching the superscript of $\Q_S^{\rm rad}$ can be dropped because with the conventional choice of dressing field its contribution cancels in the difference: 
\be
\Q^{\rm dr}_{S+} - \xoverline \Q^{\rm dr}_{S-}=0.
\ee
This shows the full agreement of \eqref{strong} and \eqref{Qmass} with \cite{Strominger_photon,Strominger_QED,Strominger_even}. The above conservation laws in even number of dimensions for massive as well as massless particles have been shown to be equivalent to the Weinberg theorem. Since a different gauge fixing is used here, we give the derivation in $A_0=0$ gauge in appendix \ref{app:soft}.

\subsection{An example}

In \cite{Susskind} a simple recipe for a finite distance construction of asymptotic charges was proposed in the example of a neutral particle decaying into charged particles. We conclude this section by explaining why it works in massless QED while it fails in massive QED and in higher than four spacetime dimensions. 

Take the Maxwell field to be zero at the initial time and suppose there is no incoming radiation. A neutral particle is sitting at rest at $\r =0$ until $t=0$, then it decays into massless charged particles. Integrating the Gauss's law,
\be\label{gauss}
\nabla\cdot \bsb E =  J^0,
\ee
along the world-volume of the sphere we obtain (using $\bsb E = \d_0{\bsb A}$ in the temporal gauge)
\be
\left. \d_\rho(\rho^2 A_\rho)+ D^a A_a\right|_{t\to \infty} = 
\sum_{k} 
Q_k \delta^2(\hat \r-\hat\r_k).
\ee
It was argued in \cite{Susskind} that the contribution of $A_\rho$ vanishes asymptotically and hence what remains coincides with the expression for the $4d$ conserved charge obtained in \cite{Strominger_photon}:
\be\label{Qlenny}
\left.\int d\hat\r \alpha(\hat\r) D^a A_a\right|_{t\to \infty}=
\sum_{k} Q_k \alpha(\hat\r_k).
\ee
In the case of massive charges this proposal gives 
\be
\sum_{k} 
\beta_k^{-1} Q_k\alpha(z_k^a),
\ee
for the r.h.s. of \eqref{Qlenny} which does not seem to agree with \eqref{Qmass} and \cite{Strominger_QED}.

The reason why this argument works in the massless case is because the retarded field of massless particles produced at $r=0$ is tangential to the sphere $R$ except possibly for a spherically symmetric divergenceless radial component. As an (unrealistic) example consider a $Z$ boson at the origin that decays at $t=t_1$ into a $W^+$ staying at rest, a massless electron that flies in $\hat \n_1$ direction and a $\bar \nu_e$. At time $t_2$ the massive boson decays into an $e^+$ escaping along $\hat \n_2$ and a $\nu_e$. The electric field of this system at a distant point looks like figure \ref{decay} and is given by
\be
\begin{split}
\bsb E(t,\r) = &\frac{e(\hat \n_1 - \hat \r \hat \r\cdot \hat \n_1)}{4\pi r(1-\hat \r\cdot \hat \n_1)}\delta(t-r-t_1)
-\frac{e(\hat \n_2 - \hat \r \hat \r\cdot \hat \n_2)}{4\pi r(1-\hat \r\cdot \hat \n_2)}\delta(t-r-t_2)\\[10pt]
&+\frac{e \hat \r}{4\pi r^2}\theta(t-r-t_1)\theta(t_2+r-t).
\end{split}
\ee
It is seen that although $A_r|_{t\to\infty} = Q(t_2-t-1)/4\pi r^2$ is nonzero, it is divergenceless and hence the approximation \eqref{Qlenny} is valid. 

\begin{figure}[t]
\centering
\includegraphics[scale = 0.5]{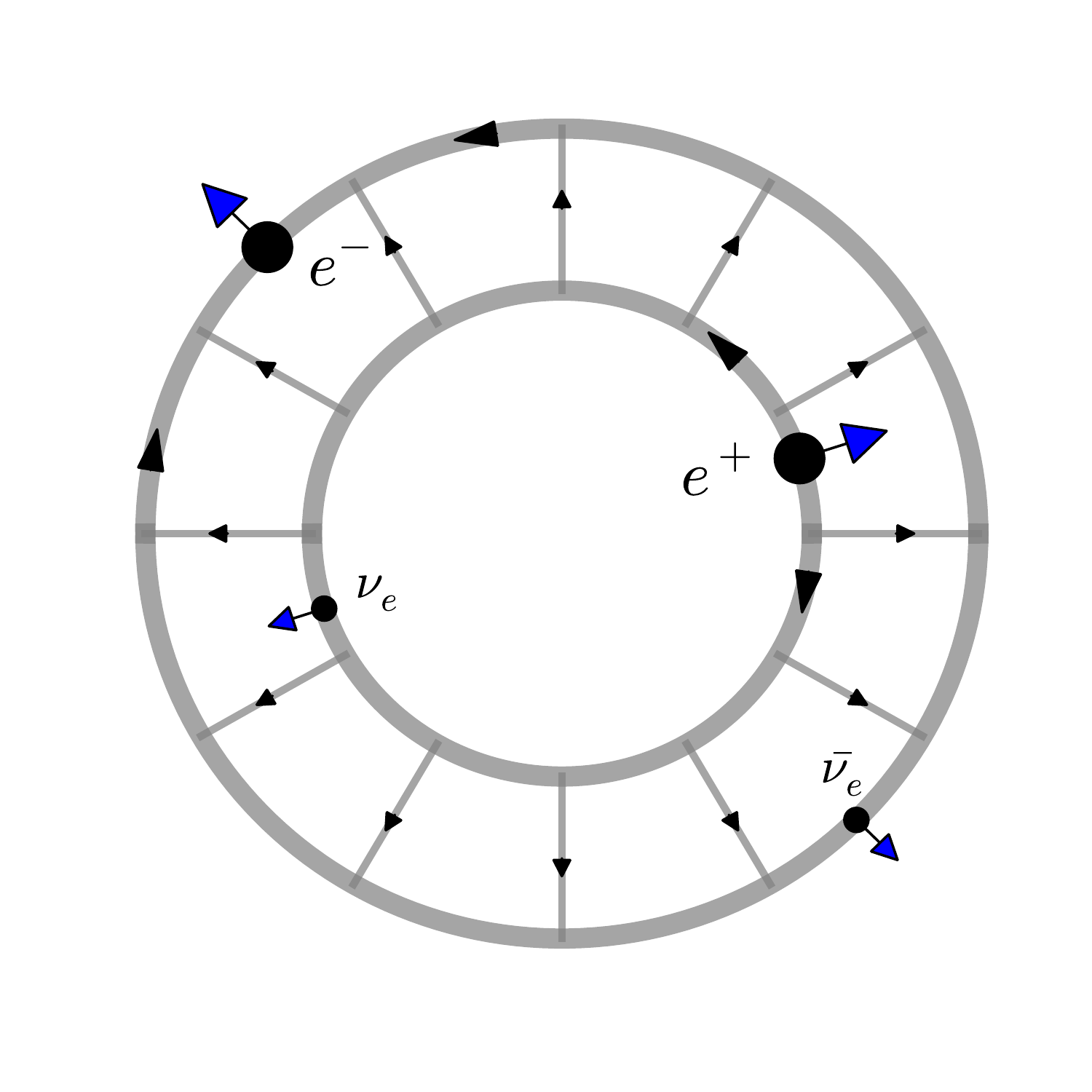} 
\caption{\small{Electric field of a pair of massless electron and positron resulting from a two step decay of a $Z$ boson at the origin.}}\label{decay}
\end{figure}

For massive charged particles, on the other hand, the radial component of the electric field is nontrivial. For instance, a slowly moving massive charged particle which crosses the sphere of radius $R$ has a field
\be
\bsb E\simeq \frac{e (\r - \bsb R)}{4\pi |\r - \bsb R|^{3/2}},
\ee
for which 
\be
\rho^{-2}\d_\rho (\rho^2 E_\rho) \simeq \frac{1}{2}\rho^{-2}D^a E_a.
\ee
This leads to a radial $A_\rho|_{t\to \infty}$ which cannot be neglected. 

In higher spacetime dimensions $A_\rho$ cannot be neglected even if the charges are massless. In that case the charge current appears at a higher order in $1/R$ expansion than the radiation field and one needs to keep the sub-leading terms in the large $R$ expansion of $A_\rho$. These are nontrivial and related to the radiative modes (see \cite{Strominger_even} and appendix \ref{app:soft}).

\section{Conclusions}

We defined adiabatic modes in electrodynamics and gravity in asymptotically flat spacetime as the subclass of large gauge transformations that can be obtained from the infinite wavelength limit of physical modes. We used the conservation of the currents associated to these symmetries to derive Weinberg theorems for scattering of soft photons and gravitons. Thus one can interpret the soft theorems as Ward identities for spontaneously broken large gauge transformations. The Minkowski vacuum transforms under these symmetries and hence the adiabatic modes are the corresponding Nambu-Goldstone modes. We derived the Ward identities for local correlation functions as well to make a closer connection with the cosmological consistency conditions. 

We also showed that the recently studied Ward identities corresponding to the asymptotic symmetry group of QED can be continued to finite distance and be obtained from the conservation of the currents associated to the adiabatic modes. There is a major difference which is that adiabatic modes correspond to a different class of large gauge transformations than those employed in \cite{Strominger_photon,Strominger_QED,Strominger_even}. Nevertheless, our analysis shows a closer connection with the cosmological consistency conditions, and offers a derivation of the antipodal matching between future and past null infinities.  We expect the generalization to the gravitational case studied in \cite{Strominger_graviton1,Strominger_graviton,Strominger_memory,Strominger_greven,Penna} be straightforward. 

Although the implications of large diffeomorphisms associated to the adiabatic modes both on correlation functions and the structure of the wavefunction of the universe are well studied, e.g. in \cite{Hinterbichler2,Pimentel,Hui,Berezhiani,Berezhiani_nontrivial,Joyce,double}, it would be interesting to understand if there is a closer analog of BMS symmetries in curved backgrounds and in cosmology. Some qualitatively different approaches can be found in \cite{Anninos,Ashtekar,Riotto}.

\section*{Acknowledgments}

We thank Lasha Berezhiani, Paolo Creminelli, Sergei Dubovsky, Victor Gorbenko, Kurt Hinterbichler, Juan Maldacena, Matias Zaldarriaga and Sasha Zhiboedov for useful discussions and comments. M.M. is supported by NSF Grants PHY-1314311 and PHY-0855425. M.S. is supported by the Institute for Advanced Study and the Raymond and Beverly Sackler Foundation.

\appendix

\section{Electromagnetic adiabatic modes in Coulomb gauge} \label{app:Noether}

In this appendix we first find adiabatic modes in Coulomb gauge as a gauge condition that fully fixes local gauge degrees of freedom. Then we give a detailed derivation of the conserved Noether current in a simple case. 

We start by imposing the Coulomb gauge condition
\be
\d_i A_i =0.
\ee
This fully fixes local $U(1)$ gauge transformations. $A_0$ is a constraint variable that can be solved from its equation of motion. However, there are residual gauge transformations with a parameter $\alpha$ which is a harmonic function with arbitrary time-dependence 
\be 
\nabla^2 \alpha(t,\x) =0.
\ee
Such gauge transformations are necessarily nontrivial at spatial infinity. They transform the vacuum into a state with nontrivial $A_\mu$ and hence, by virtue of being a symmetry, generate new solutions of the theory. These solutions are unphysical since they have infinite wavelength. However, there is a subclass of them with time-independent $\alpha$ that can be continued to finite wavelength: The homogeneous constraint equation
\be
\nabla^2 A_0 = 0
\ee
with $A_0 = \d_0\alpha$ is satisfied for any harmonic $\alpha(t,\x)$. Once we deform this solution to finite wavelength $\nabla^2\to -\q^2\neq 0$. Thus in order to ensure continuity we make the stronger requirement that $A_0(\q =0)=0$. This implies
\be
\d_0\alpha =0.
\ee
This subclass of large gauge transformations generate the adiabatic modes. Note that since the dynamical source-free equation for $A_i$ is of the form $(-\d_0^2 +\nabla^2)A_i=0$ there is no obstacle in continuing an infinite wavelength solution $A_i(\q =0)$ to finite wavelength. $\d_0^2A_i$ can adjust itself to cancel $\nabla^2 A_i$. 

\subsection{Noether current}

Next we derive the Noether current. The equation of motion for $A_i$ after solving for $A_0$ from the constraint $\nabla^2A_0 =-J_0$ and using $\d_\mu J^\mu =0$ reads
\be\label{Aia}
\Box A_i =(-\d_0^2 +\nabla^2)A_i=- J^T_i \equiv -J_i + \frac{\d_i\d_j}{\nabla^2}J_j.
\ee
The Noether current for $\alpha$ transformation is
\be
K^\mu = \d_i \alpha \d^\mu A_i - \d^\mu\d_i\alpha A_i + \alpha J^\mu
-\delta^\mu_i \d_i \alpha \frac{\d_j}{\nabla^2}J_j,
\ee
whose conservation can be verified using \eqref{Aia}. Unlike temporal gauge this current has no redundancy, it is conserved only for adiabatic modes. The last term ensures transversality of the source but it won't play any role in the derivation of soft theorems.

Let us derive the first few terms in $K^\mu$ by varying the action in a simple example. Consider the action for photon plus a charged field
\be
S=\int -\frac{1}{4} F_{\mu\nu}^2 - |(\d_\mu -iQ A_\mu)\phi|^2.
\ee
In Coulomb gauge the Maxwell term becomes
\be
S_C = \int -\frac{1}{2} (\d_\mu A_i)^2 - A_0 \nabla^2 A_0 .
\ee
implying that $A_0$ is a constraint variable. The equation of motion for $A_0$ is
\be
\nabla^2 A_0 = iQ (\phi^\dagger \d_0 \phi - \d_0\phi^\dagger \phi - 2iQ A_0 \phi^\dagger \phi).
\ee
Hence $A_0$ starts quadratic in the fields and does not contribute to the quadratic and cubic action:
\be
S_{C} = \int -\frac{1}{2} (\d_\mu A_i)^2 + |\d_0\phi|^2 -|\d_i\phi|^2 + A_i J_i^T+\cdots
\ee
where we defined the electric current 
\be
J_\mu = iQ (\phi\d_\mu \phi^\dagger-\phi^\dagger \d_\mu \phi ),
\ee
and $J^T_i$ is its transverse spatial component
\be
J_i^T = J_i - \frac{\d_i\d_j}{\nabla^2} J_j
\ee
satisfying $\d_i J_i^T=0$. Given that $A_i$ is constrained to be transverse it is important to cancel its coupling to the longitudinal component of the current in order to get correct equation of motion for $A_i$. Keeping those terms would have corresponded to introducing a longitudinal component for $A_i$ when varying the action.

Up to boundary terms this action is invariant under the transformation 
\be
A_i \to A_i +\d_i \alpha ,\qquad \phi \to (1+iQ \alpha) \phi
\ee
if $\d_0\alpha = \nabla^2 \alpha =0$. This is a subgroup of large gauge transformations. To see this note that
\be
\begin{split}
\delta\ \frac{1}{2}(\d_\mu A_i)^2 =~ & - A_i (-\d_0^2 +\nabla^2) \d_i \alpha +\d_\mu(A_i\d^\mu\d_i\alpha),\\[10pt]
\delta\ |\d_0\phi|^2 =~ & 0,\\[10pt]
\delta\ |\d_i\phi|^2 =~ & iQ \d_i \alpha (\phi \d_i\phi^\dagger - \phi^\dagger \d_i\phi) = \d_i \alpha \ J_i, \\[10pt]
\delta\ A_i J_i^T =~ & \d_i \alpha \ J_i + \nabla^2 \alpha \frac{\d_j}{\nabla^2}J_j 
-\d_i\left(\d_i\alpha \frac{\d_j}{\nabla^2}J_j\right)+\text{cubic}.
\end{split}
\ee
To derive the Noether current, modify the field variations to 
\be
A_i \to A_i +\ep \d_i \alpha ,\qquad \phi \to (1+iQ \ep\alpha) \phi,
\ee
with $\ep$ an arbitrary spacetime dependent function. There will be new terms in the variation of the action
\be
\begin{split}
\delta\ \frac{1}{2}(\d_\mu A_i)^2 =~ & \d_\mu\ep \d_i \alpha \d_\mu A_i -\d_\mu\ep \d_\mu\d_i \alpha A_i 
+\d_\mu(\ep A_i\d^\mu\d_i\alpha),\\[10pt]
\delta\ |\d_0\phi|^2 =~ & iQ\alpha \d_0 \ep (\phi \d_0\phi^\dagger - \phi^\dagger \d_0\phi),\\[10pt]
\delta\ |\d_i\phi|^2 =~ & iQ \d_i(\ep\alpha) (\phi \d_i\phi^\dagger - \phi^\dagger \d_i\phi),\\[10pt]
\delta\ A_i J_i^T =~ &\ep \d_i \alpha J_i + \d_i \ep \d_i \alpha \frac{\d_j}{\nabla^2}J_j ,
-\d_i\left(\ep \d_i\alpha \frac{\d_j}{\nabla^2}J_j\right)+\text{cubic}.
\end{split}
\ee
Thus the total variation of the action is of the form
\be
\delta S = -\int \d_\mu \ep \ K^\mu,
\ee
where 
\be
K^\mu = \d_i \alpha \ \d^\mu A_i - \d^\mu\d_i\alpha \ A_i +\alpha J^\mu 
- \delta^\mu_i \d_i\alpha \frac{\d_j}{\nabla^2}J_j.
\ee
On-shell, $\delta S$ must vanish up to surface terms for arbitrary $\ep(x)$, hence $\d_\mu K^\mu =0$.}

\section{LSZ reduction and local Ward identities}\label{app:LSZ}

In this appendix we show that 
\be\label{LSZ}
\begin{split}
\lim_{q\to 0}\int d^4x e^{-iq\cdot x} &\d_\mu\expect{\hat T\{K^\mu(x) \phi(x_1)\cdots\phi(x_N)\}}\\[10pt]
= \sum_n \frac{(i)^{n+1}}{n!}\bvep_{i_0\cdots i_n}&\frac{\d^{n}}{\d \q_i^{i_1}\cdots \d \q^{i_n}}
\langle A_{i_0}(q)|\hat T\{\phi(x_1)\cdots \phi(x_N)\}|0\rangle ,
\end{split}
\ee
where $q$ is a soft positive-frequency null four-vector. To see this first integrate the l.h.s. by parts
\be\label{qK}
\begin{split}
i\int e^{-iq\cdot x} q_\mu\expect{\hat T\{K^\mu(x) \phi(x_1)\cdots\phi(x_N)\}}&\\[10pt]
+ e^{iq^0T(1-i\ep)}\expect{K^0(T,\q) \hat T\{\phi(x_1)\cdots\phi(x_N)\}}&
-e^{-iq^0T(1-i\ep)}\expect{ \hat T\{\phi(x_1)\cdots\phi(x_N)\}K^0(-T,\q)},
\end{split}
\ee
where $\pm T$ are the limits of time-integration which will be sent to infinity. We restored the $i\ep$ prescription. For $q^0>0$ the boundary term at $t=-T$ can be dropped. This is because the positive frequency modes of any quantum field that appears in $K^0(-T,\q)$,  which are of the form
\be
a e^{iE_\p T(1-i\ep)} ,\qquad E_\p >0,
\ee
annihilate the in-vacuum. And the negative frequency modes which create particles vanish by the $i\ep$ prescription. Moreover, the $i\ep$ prescription kills all non-linear terms in the boundary term at $t =T$. Consider some quadratic operator $\phi^2(T,\q)$ appearing inside $K^0$. The non-vanishing part is when we pick two positive frequency modes which create particles acting on out-vacuum:
\be
a_{\p_1} a_{\p_2} e^{-i(E_{\p_1}+E_{\p_2}) T(1-i\ep)},\qquad \p_1+\p_2 = \q.
\ee
However, the momentum condition enforces $E_{\p_1}+E_{\p_2}\geq q^0$ and this term goes to zero exponentially as we take $T\to \infty$, except for the uninteresting colinear case $\p_1\parallel \p_2$. What survives is the soft term in $K^0( T,\q)$,
\be\label{K^0}
K^0_S(T,\q) =-\int d^3\x e^{-i\q\cdot\x} \d_i\alpha(\x) \d_0 A_i(T,\x),
\ee
which generates an outgoing photon. Let us first take $\d_i\alpha$ outside of the integral to obtain a differential operator that projects the soft photon onto a specific configuration. Using the expansion \eqref{vep}
\be
K^0_S(T,\q) =-\sum_n \frac{(i)^{n}}{n!}\bvep_{ii_1\cdots i_n}\frac{\d^{n}}{\d \q_i^{i_1}\cdots \d \q^{i_n}}
\int d^3\x e^{-i\q\cdot\x} \d_0 A_i(T,\x).
\ee
Next plug in the mode expansion \eqref{modes} to get 
\be\label{1photon}
\langle 0|\d_0 A_i(T,\q) e^{iq^0T(1-i\ep)} = -\frac{i}{2}\sum_s \langle 0|a(s,\q)\vep_i^*(s,\q).
\ee
This gives $1/2$ of the r.h.s. of \eqref{LSZ}. 

Finally, we should consider the first term in \eqref{qK}. In the $q\to 0$ limit all but the soft term in $K^\mu$ can be neglected. This is because all operator insertions are at finite positions and therefore all propagators are off-shell. Hence unlike in derivation of Weinberg theorem the insertion of quadratic and higher order terms of $K^\mu$ as interaction vertices will be regular in the limit $q\to 0$. However, the soft term in $K^0$ leads to an LSZ pole:
\be
\begin{split}
-i\int d^4x e^{-iq\cdot x} q^0\expect{\hat T\{\d_i\alpha(x)\d_0A_i(x) \phi(x_1)\cdots\phi(x_N)\}}&\\[10pt]
=-iq^0\int d^3\x e^{-i\q\cdot \x} e^{iq^0T(1-i\ep)}
\d_i \alpha(\x)\expect{A_i(T,\q) \hat T\{\phi(x_1)\cdots\phi(x_N)\}}&
+\O(q^2),
\end{split}
\ee
where we used the fact that $\d_0\alpha(\x) =0$, and dropped the boundary term at $-T$ for the same reasons discussed above. We also dropped the contact terms arising from taking $\d_0$ outside the time-ordered product since they contain equal-time commutators $[A_i(t,\x),\phi(t,\x')]=0$. This gives an identical contribution as \eqref{K^0}. Together, they reproduce the r.h.s. of \eqref{LSZ}.

\section{A freely moving charge and its dressing field}\label{app:dress}

In this appendix we will verify that the asymptotic charge $\Q$ vanishes for a freely moving massless charge. This is a trivial statement, given that $\Q$ is the integral of a total derivative (an exact form) over a closed surface. So the real goal of this simple example is to see when and at what rate various contributions from the electric current and the dressing field of the charge are deposited as it moves through the sphere. The most straightforward way to see this is to use Maxwell equations to write 
\be\label{Qfree}
\Q =  R^{d-2}\int d\hat\r\alpha(\hat\r) \int_{-T}^{T} dt (- D^aF_{\rho a}  +J_\rho)
= R^{d-2}\int d\hat\r\alpha(\hat\r) \int_{-T}^{T} dt \ \d_0 E_r,
\ee
where we used the fact that massless particles enter the sphere well after $-T$ and leave well before $T$ and hence the space-like surfaces at $\pm T$ can be neglected. 

\begin{figure}[t]
\centering
\includegraphics[scale = 0.5]{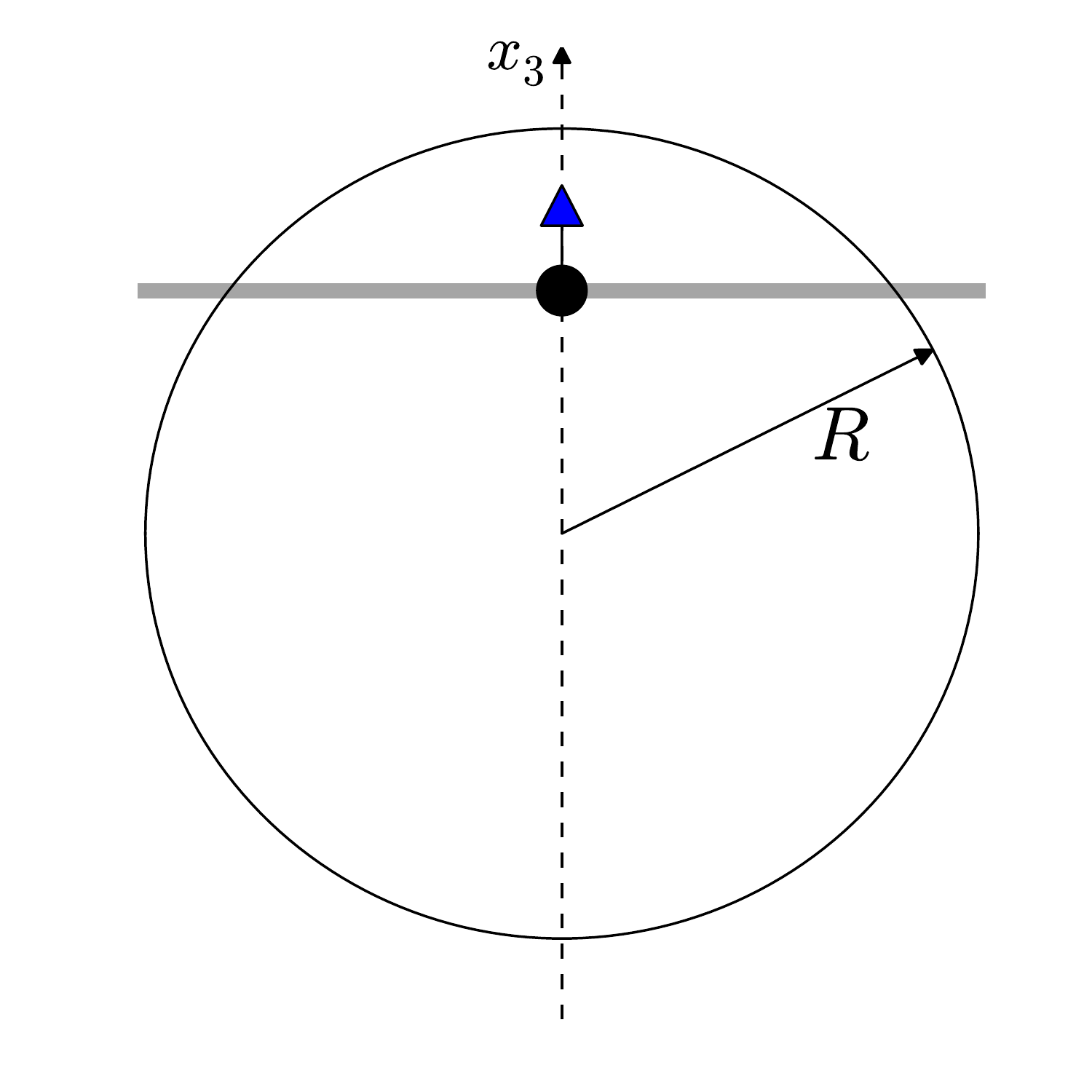} 
\caption{\small{A massless charge moving inside the sphere $R$ along the third axis. The electric field is confined in a plane.}}\label{free}
\end{figure}

Let us restrict to $4d$ for simplicity. There is no subtlety in continuation to higher dimensions. A massless charge $Q$ moving along $x_3$ has an electric field which is confined in a plane perpendicular to $\hat \x_3$, given by
\be\label{Eperp}
\E_\perp = \frac{Q}{4\pi r_\perp}\delta(x_3-t),
\ee
where $r_\perp =\sqrt{x_1^2+x_2^2}$. $\Q=0$ because $\E_r(\pm T,R,\hat \r)=0$ for all $\hat \r$. To understand how this happens in a more local sense let us choose $\alpha(\hat\r)$ to peak in a specific direction $\hat \r_0 =(\theta_0,\vphi_0)$. Then the only time the integrand in \eqref{Qfree} is nonzero is at $t_0 = R\cos\theta_0$, see figure \ref{free}. And it integrates to zero since $\E_r$ is zero immediately before and immediately after the plane crosses at $t_0$. 

Let us also consider the case where $\alpha$ peaks at $\hat \r_0 = \hat \x_3$ where the charge exits the sphere. The relevant time is $t_0 =R$ and therefore we can use 
\be\label{Q4}
\Q = \int d\hat\r \alpha(\hat \r) \int_{-R}^{T} du (\d_u D^aA_a-\d_r D^aA_a + D^2 A_r - R^2 J_u).
\ee
At this point we can write $u= t-x_3$, $r=x_3$, $\A_\perp =r^{-1} \A_a$. For the field \eqref{Eperp} we have in $A_0 =0$ gauge
\be
\A_\perp =- \frac{Q}{4\pi r_\perp}\theta(u),
\ee
implying that 
\be\label{drAa}
\d_r \A_a = \A_r =0.
\ee
We can also use the fact that the divergence of $\E$ is purely within the plane, the Gauss law, and that $\d_u=\d_0$ and $J_u = J_t$ to write
\be
-J_u = \nabla\cdot \E = \nabla_\perp \cdot \E = -R^{-2} \d_u D^a\A_a.
\ee
This together with \eqref{drAa} implies that the integrand of \eqref{Q4} is zero. Hence the contributions from the field and the electric current instantaneously cancel.

\section{Soft photon theorems from BMS charges}\label{app:soft}

In this appendix we show the equivalence of the conservation laws derived in section \ref{sec:BMS} and the Weinberg theorem in even dimensional massless and massive QED. Despite a slightly different gauge choice the derivation is almost identical to those given in \cite{Strominger_photon,Strominger_QED,Strominger_even}. The only new part is a simple generalization to the massive QED in higher than $4d$ based on a conjectured differential identity.

We consider a $2m+2$ dimensional spacetime with radial metric
\be
ds^2 = -dt^2 +d\rho^2 +\rho^2 \gamma_{ab} dz^a dz^b
\ee
where $\gamma_{ab}$ is the metric of a unit $2m$-sphere. The Maxwell equation reads
\be\label{Max}
\d_\mu(\sqrt{-g} g^{\mu\nu}g^{\alpha\beta}F_{\nu\beta})=\sqrt{-g} J^\alpha.
\ee
We fix the temporal gauge $A_0=0$ together with $A_\rho|_{t=-\infty} =0$. For the late radiation we use retarded coordinates 
\be
u=t-\rho,\qquad r =\rho.
\ee
The components of the gauge field and derivatives are related to those in $t,r$ coordinates according to
\be
A_u = A_0 = 0,\qquad A_r= A_\rho,\qquad \d_\rho = \d_r -\d_u,\qquad \d_0 =\d_u.
\ee
The $t$ and $\rho$ components of \eqref{Max} give
\be\label{t}
-\d_u^2 A_r +r^{-2} \d_u D^aA_a +r^{-2m}\d_r(r^{2m}\d_u A_r)=J^t,
\ee
\be\label{r2}
-\d_u^2 A_r +r^{-2} \d_u D^a A_a + r^{-2} (D^2 A_r -\d_r D^a A_a)=J^r,
\ee
where $D_a$ is the covariant derivative with respect to $\gamma_{ab}$. Similar equation can be derived for early radiation in advanced coordinates. 

The conservation law is obtained by integrating the conserved current over a closed surface consisting of the world-volume of a sphere of radius $R$ capped by two spatial patches at $\pm T$. As argued in section \ref{sec:qed} the contribution of the dressing fields effectively gives an antipodal matching. Hence for scattering of massless in and out charged particles, we have 
\be\label{Q}
\begin{split}
\Q_{S+}-\xoverline \Q_{S-} =& R^{2m-2}\int d\hat \r \alpha(\hat \r)\\[10pt]
\out\Big[\int_{-R}^{T} du &(\d_u D^aA_a-\d_r D^aA_a + D^2 A_r)
-\int_{-T}^{R} dv (\d_v D^aA_a+\d_r D^aA_a - D^2 A_r)_{-\hat \r}\Big]\inn\\[10pt]
=&  \sum_{k} \eta_k Q_k \alpha(\hat \r_k) \out\innn\rangle
\end{split}
\ee
Only the radiation field contributes to the above difference of soft charges. To derive the Weinberg's theorem from these conservation laws, we need to express the gauge fields appearing in $\Q_S$ in terms of the radiation data. The radiation field scales as $A_a = \O(r^{1-m})$, while because of the overall $R^{2m-2}$ \eqref{Q} is sensitive to $A_a^{(2m-2)},A_a^{(2m-3)}$, and $A_r^{(2m-2)}$ where we adopted the expansion 
\be
A= \sum_{n=0}^\infty \frac{A^{(n)}}{r^n}.
\ee
Naively, at $m>1$ there are contributions to \eqref{Q} that are divergent in the $R\to \infty$ limit. However, their vanishing is an automatic consequence of homogeneous Maxwell equations. The $O(R^0)$ term is the one that we are interested in because it relates electric flux to soft radiation. At $m>1$ the relevant components of soft field, namely $A_a^{(2m-2)},A_a^{(2m-3)}$, and $A_r^{(2m-2)}$ are sub-leading in $R$-scaling compared to the radiative modes. For $m=1$ corresponding to $4d$ the radiation field appears at the same level. We first express the radiation field $A_a^{m-1}$ in terms of the creation and annihilation operators of the free asymptotic fields $A^{\rm out/in}_\mu$, and then treat $4d$ and higher dimensional spacetimes separately. 

\subsection{Radiative modes}

For the out-field we start from the free field expansion
\be\label{Amu}
A_\mu(t,\r) = \sum_s \int \frac{d^{2m+1}\q}{(2\pi)^{2m+1} 2\omega}[\vep_\mu^*(s,\hat q)a(s,\q) 
e^{-i\omega t+i\q\cdot \r}+\text{c.c.}],
\ee
and use the fact that in spherical coordinates
\be\label{AaAi}
A_a = \d_a x^i A_i = r \d_a\hat \r^i A_i.
\ee
In the large $r$ limit the integration over $\hat \q$ can be performed using saddle point approximation. There is one stationary point at $\hat q=\hat \r$ around which we can expand $\hat q\cdot \hat \r= 1-\frac{1}{2}\sum_{i=1}^{2m} w_i^2$, where $w_i$ parameterize $2m$ angular directions, and another one at $\hat q = -\hat \r$. The result for the latter is proportional to $e^{2i\omega r}$ and hence it is negligible in $r\to \infty$ limit. The first saddle gives
\be
e^{-i\omega t}\int d^{2m}\hat q \vep_i^*(s,\hat q) a(s,\q)e^{i\omega r\hat q\cdot \hat \r}
= e^{-i\omega u} \vep_i^*(s,\hat \r) a(s,\omega \hat \r)\left(\frac{2\pi}{i\omega r}\right)^m.
\ee
Substituting in \eqref{Amu} using \eqref{AaAi} and taking the Fourier transform with respect to $u$ gives
\be\label{Aa}
A_a^{(m-1)}(\omega_0,\hat\r)=\int_0^\infty d\omega \frac{(-i\omega)^m}{(2\pi)^m 2\omega}\d_a\hat \r^i
\sum_s[\vep_i^*(s,\hat \r) a(s,\omega \hat \r) \delta(\omega-\omega_0)+(-1)^{m}
\vep_i(s,\hat \r) a^\dagger(s,\omega \hat \r)\delta(\omega+\omega_0)].
\ee
Repeating the same procedure for the in-field, we get the dominant saddle point at $\hat \q = -\hat \r$. The creation and annihilation operators are respectively proportional to $e^{-i\omega v}$ and $e^{i\omega v}$. Taking the Fourier transform of the in-field with respect to $v$ we get
\be\label{Aav}
A_a^{(m-1)}(\omega_0,-\hat\r)=-\int_0^\infty d\omega \frac{(-i\omega)^m}{(2\pi)^m 2\omega}\d_a\hat \r^i
\sum_s[(-1)^{m}\vep_i^*(s,\hat \r) a(s,\omega \hat \r) \delta(\omega-\omega_0)+
\vep_i(s,\hat \r) a^\dagger(s,\omega \hat \r)\delta(\omega+\omega_0)].
\ee

\subsection{Four dimensions}

Dropping the angular integral, the soft piece of the current is given, up to terms proportional to negative powers of $R$, by
\be\label{4dQ}
\int_{-R}^T du [\d_u D^a A_a^{(0)}+ D^2 A_r^{(0)}-D^a A_a^{(-1)}]
-\int_{-T}^R dv [\d_v D^a A_a^{(0)}- D^2 A_r^{(0)}+D^a A_a^{(-1)}]_{-\hat\r}.
\ee
The last two terms in both brackets are zero, and for the first we use the mode expansion. For a finite sphere of radius $R$ the limits of the $u$ ($v$) integration are finite but large. The difference compared to the asymptotic case considered in \cite{Strominger_photon}, where the integration is over $-\infty<u<\infty$ and $\omega_0 =0$ in \eqref{Aa}, is that here the delta function becomes smooth and instead of just the zero mode all modes with $\omega R\ll 1$ contribute. 

In order to use \eqref{Aav}, either for $\omega R\ll 1$ or the zero mode as in \cite{Strominger_photon}, one has to make a continuity assumption since the saddle point approximation made in the derivation of \eqref{Aav} is valid for $\omega r \gg 1$. If the scattering process has a finite duration $\tau\ll R$, we expect the integrals in \eqref{4dQ} to converge very fast. Hence one could take the integral over a shorter period $ \tau\ll \Delta T\ll R$ to pick up contribution from all modes with $\omega \Delta T\ll 1$. Most of these satisfy $\omega R\gg 1$ for which \eqref{Aav} holds.

Since the integral in \eqref{Aa} is on positive $\omega$ we get half of the contribution from each delta function, so the outgoing radiation contributes 
\be\label{QS4}
\Q_{S+}=\frac{1}{2} \lim_{\omega \to 0} \frac{-1 }{4\pi} D^a \d_a \hat \r^i\sum_s 
[\vep_i^*(s,\hat \r)\omega  a_{\rm out}(s,\omega \hat \r)+
\vep_i(s,\hat \r)\omega a_{\rm out}^\dagger(s,\omega \hat \r)].
\ee
Since $\Q_{S+}$ is long after scattering and we are working in the conventional boundary condition with no radiation in the in- and out-states the part proportional to $a^\dagger(s,\q)$ can be dropped. We also need to subtract the incoming soft contribution
\be
\xoverline \Q_{S-}=\frac{1}{2} \lim_{\omega \to 0} \frac{-1 }{4\pi} D^a \d_a \hat \r^i\sum_s 
[\vep_i^*(s,\hat \r)\omega  a_{\rm in}(s,\omega \hat \r)+
\vep_i(s,\hat \r)\omega a_{\rm in}^\dagger(s,\omega \hat \r)].
\ee
however after using
\be
a(s,\q) \S = - \S a^\dagger(s,\q)
\ee
derived in section \ref{sec:anti}, we get the same contribution as $\Q_{S+}$. So we just multiply it by $2$. Also as argued in \ref{sec:anti} only one linear combination of polarization vectors is coupled hence to show the equivalence with the Weinberg theorem we can plug in its prediction for emission amplitude:
\be\label{wein3}
\lim_{\omega\to 0} \omega \out a(s,\omega \hat \r)\inn = 
\sum_k \frac{\eta_k Q_k\vep(s,\hat \r)\cdot p_k}{(E_k-\hat \r \cdot \p_k)}
\out {\rm in}\rangle,
\ee
in $\Q_{S+}-\xoverline \Q_{S-}$ and check if it agrees with the r.h.s. of \eqref{Q}. To obtain that from this equation we apply the same operation as in \eqref{QS4}. On the r.h.s. we choose $\vep(s,\hat \r)$ to be spatial and transverse (by subtracting a piece proportional to $q_\mu$) so that
\be
\sum_s \vep^*_i(s,\hat \r)\vep_j(s,\hat \r)=\delta_{ij}- \hat \r_i \hat \r _j
\ee
to write for massless particles with $\p = E \hat p$
\be
\d_a \hat \r^i \sum_s \bvep^*(s,\hat\r) \frac{\bvep(s,\hat \r)\cdot \p}{(E-\hat \r \cdot \p)}
= -\d_a \log(1- \hat \r\cdot\hat \p).
\ee
To find the action of the spherical derivatives in \eqref{QS4} on this function, define $\cos \theta =\hat \r\cdot\hat p$ and use
\be
\sin\theta D^a\d_a \log(1- \cos \theta)= {4\pi}\delta(\hat \r) -1,
\ee
Plugging back in \eqref{QS4} and using the conservation of total charge $\sum_k \eta_k Q_k =0$ gives the r.h.s. of \eqref{Q} as desired.

\subsection{Even dimensions higher than four}

For $m>1$ we need to relate higher order terms in $1/r$ expansion of $A_r$ and $A_a$ to the radiative modes using the asymptotic form of the Maxwell equations. First, we use the fact that $J^t=J^r$ and subtract \eqref{t} from \eqref{r2} to get
\be\label{ua}
r^{-2m}\d_r(r^{2m}\d_u A_r)= r^{-2} (D^2 A_r -\d_r D^a A_a).
\ee
This, in particular, implies that
\be\label{ua1}
D^aA_a^{(2m-2)}=-\frac{1}{2m-2} D^2A_r^{(2m-1)}.
\ee
Combining with lower order terms in $1/R$ expansion of \eqref{t} and \eqref{r2}, and using the fact that asymptotically the electric currents start from $J^\mu\propto 1/r^{2m}$, we get
\be
\d_u^2 A_r^{(n)} = -\frac{1}{2(n-m-1)}[D^2-(2m-n+1)(n-2)]\d_uA_r^{(n-1)}
\ee
and 
\be
[D^2 -(2m-n+1)(2m-n)]\d_uA_r^{(n)} = [D^2-(2m-n+1)(n-2)]D^a A_a^{(n-2)},
\ee
which in particular implies 
\be
\d_u A_r^{(m+1)}=D^a A_a^{(m-1)}.
\ee
Using these relation we can express $Q_{S+}$ in terms of radiation data. We first use \eqref{ua} to rewrite $Q_{S+}$ as
\be
\d_u D^aA_a-\d_r D^aA_a + D^2 A_r = \d_u D^aA_a +r^{2-2m}\d_r(r^{2m}\d_u A_r).
\ee
Multiplying by $R^{2m-2}$ we get for the zeroth order term in $1/R$ expansion of $\Q_{S+}$
\be
\d_u D^aA_a^{(2m-2)}+\d_u A_r^{(2m-1)}.
\ee
Using \eqref{ua1} this becomes
\be
-\frac{1}{2(m-1)}[D^2-2(m-1)] \d_u A_r^{(2m-1)}.
\ee
Going to the Fourier space and using the recursive formulas we finally get
\be\label{duAr}
\Q_{S+} =  \frac{(-1)^{m+1}(\d_u)^{2-m}}{2^{m-1}\Gamma(m)}
\prod_{n=2m}^{m+2}[D^2-(2m-n+1)(n-2)] D^a A_a^{(m-1)}(\omega = 0),
\ee
Substituting \eqref{Aa} and using the same Hermitian definition of $\omega =0$ as in the $4d$ case gives 
\be\label{QS}
\begin{split}
\Q_{S+}=\frac{1}{2}\lim_{\omega \to 0^+} \int d\hat\r \alpha(\hat\r) \frac{(-1)^m }{2^m (2\pi)^m \Gamma(m)}&\\[10pt]
\prod_{n=2m}^{m+2}[D^2-(2m-n+1)(n-2)] 
D^a \d_a \hat \r^i\sum_s &
[\vep_i^*(s,\hat \r)\omega  a_{\rm out}(s,\omega \hat \r)+
\vep_i(s,\hat \r)\omega a_{\rm out}^\dagger(s,\omega \hat \r)].
\end{split}
\ee
The part containing creation operators annihilates the out-state, but we need to subtract $\xoverline \Q_{S-}$ whose contribution is the same as $-\Q_{S+}$. So we just cancel the factor $1/2$ in the above expression. To check with the Weinberg theorem we apply the same operation as in \eqref{QS} to \eqref{wein3} to obtain $\Q_S$ on the l.h.s. On the r.h.s. we use
\be
\sum_s \vep^*_i(s,\hat \r)\vep_j(s,\hat \r)=\delta_{ij}- \hat \r_i \hat \r _j
\ee
to write
\be
\d_a \hat \r^i \sum_s \bvep^*(s,\hat\r) \frac{\vep(s,\hat \r)\cdot p_k}{E_k(1-\hat \r \cdot \hat p_k)}
= -\d_a \log(1- \hat \r\cdot\hat p).
\ee
To find the action of the spherical derivatives in \eqref{QS} on this function, define $\cos \theta =\hat \r\cdot\hat p$ and use
\be
D^a\d_a \log(1- \cos \theta)= \frac{2(m-1)}{1-\cos \theta}-2m+1,
\ee
and 
\be
D^2 \frac{1}{(1-\cos\theta)^n} = -\frac{2n(m - l-1)}{(1-\cos\theta)^{n+1}}+\frac{n(2m-n-1)}{(1-\cos\theta)^n}.
\ee
Using the conservation of total charge $\sum_k \eta_k Q_k =0$, and the expression for the area of $(2m-1)$-sphere $2\pi^m/\Gamma(m)$ we obtain
\be
\begin{split}
-\prod_{n=2m}^{m+2}&[D^2-(2m-n+1)(n-2)] D^a\d_a \sum_k \eta_k Q_k\log(1- \hat \r\cdot\hat p_k)\\[10pt]
&= (-1)^{m} \Gamma(m)2^m (2\pi)^m \sum_{k} \eta_k Q_k \delta(\hat\r-\hat \r_k) 
\end{split}
\ee
which after including the overall factor in \eqref{QS} gives the desired result. 

\subsection{Massive QED}

As seen in section \ref{sec:massive} in the presence of massive in and out states the r.h.s. of the asymptotic conservation law \eqref{Q} is modified by the following sum over massive charges 
\be\label{QHmass}
\int d\hat\r \alpha(\hat\r)\sum_{k} 
\frac{\eta_k Q_k}{\Omega_{2m}\gamma_k^{2m}(1-\hat\r\cdot\bsb \beta_k)^{2m}}
\ee
To see that this piece agrees with the Weinberg theorem apply the operator in the expression \eqref{QS} for the soft charge to terms with massive particles on the r.h.s. of the soft theorem \eqref{wein3}. We have
\be
\d_a \hat \r^i \sum_s \bvep^*(s,\hat\r) \frac{\vep(s,\hat \r)\cdot p_k}{(E_k-\hat \r \cdot \p_k)}
= -\d_a \log(1- \hat \r\cdot\bsb \beta).
\ee
Defining $\hat \r\cdot\bsb \beta = \beta \cos \theta$ and using the following relation, that holds for the first few values of $m$ and we conjecture that it holds identically, 
\be
\prod_{n=2m}^{m+2}[D^2-(2m-n+1)(n-2)]D^a \d_a \log(1-\beta \cos\theta)
= \frac{(-1)^m \Gamma(2m)}{\gamma^{2m}(1-\beta\cos\theta)^{2m}}
\ee
and $\Omega_{2m} = (4\pi)^m \Gamma(m)/\Gamma(2m)$ we get \eqref{QHmass}.

\bibliography{bibwein}

\end{document}